\makeatletter
\declare@file@substitution{revtex4-1.cls}{revtex4-2.cls}
\makeatother

\documentclass[ApJ]{aastex631}
\usepackage{amsmath}

\shortauthors{A. Lapi et al.}
\shorttitle{Little Ado About Everything}

\begin{document}

\title{Little Ado about Everything: $\eta$CDM, a Cosmological Model with \\Fluctuation-driven Acceleration at Late Times}

\author[0000-0002-4882-1735]{Andrea Lapi}
\affiliation{SISSA, Via Bonomea 265, I-34136 Trieste, Italy}\affiliation{IFPU - Institute for fundamental physics of the Universe, Via Beirut 2, 34014 Trieste, Italy}\affiliation{INFN-Sezione di Trieste, via Valerio 2, 34127 Trieste,  Italy}\affiliation{INAF/IRA, Istituto di Radioastronomia, Via Piero Gobetti 101, 40129 Bologna, Italy}

\author[0000-0003-3127-922X]{Lumen Boco}
\affiliation{SISSA, Via Bonomea 265, I-34136 Trieste, Italy}\affiliation{IFPU - Institute for fundamental physics of the Universe, Via Beirut 2, 34014 Trieste, Italy}

\author[0000-0003-4537-0075]{Marcos M. Cueli}
\affiliation{SISSA, Via Bonomea 265, I-34136 Trieste, Italy}\affiliation{IFPU - Institute for fundamental physics of the Universe, Via Beirut 2, 34014 Trieste, Italy}

\author[0000-0002-9153-1258]{Balakrishna S. Haridasu}
\affiliation{SISSA, Via Bonomea 265, I-34136 Trieste, Italy}\affiliation{IFPU - Institute for fundamental physics of the Universe, Via Beirut 2, 34014 Trieste, Italy}

\author[0000-0002-3515-6801]{Tommaso Ronconi}
\affiliation{SISSA, Via Bonomea 265, I-34136 Trieste, Italy}\affiliation{IFPU - Institute for fundamental physics of the Universe, Via Beirut 2, 34014 Trieste, Italy}

\author[0000-0002-8211-1630]{Carlo Baccigalupi}
\affiliation{SISSA, Via Bonomea 265, I-34136 Trieste, Italy}\affiliation{IFPU - Institute for fundamental physics of the Universe, Via Beirut 2, 34014 Trieste, Italy}\affiliation{INFN-Sezione di Trieste, via Valerio 2, 34127 Trieste,  Italy}\affiliation{INAF, Osservatorio Astronomico di Trieste, Via G. B. Tiepolo 11, 34131 Trieste, Italy}

\author[0000-0003-1186-8430]{Luigi Danese}
\affiliation{SISSA, Via Bonomea 265, I-34136 Trieste, Italy}\affiliation{IFPU - Institute for fundamental physics of the Universe, Via Beirut 2, 34014 Trieste, Italy}

\begin{abstract}
We propose a model of the Universe (dubbed $\eta$CDM) featuring a controlled stochastic evolution of the cosmological quantities, that is meant to render the effects of small deviations from homogeneity/isotropy on scales of $30-50\, h^{-1}$ Mpc at late cosmic times, associated to the emergence of the cosmic web. Specifically, we prescribe that the behavior of the matter/radiation energy densities in different patches of the Universe with such a size can be effectively described by a stochastic version of the mass-energy evolution equation. The latter includes, besides the usual dilution due to cosmic expansion, an appropriate noise term that statistically accounts for local fluctuations due to inhomogeneities, anisotropic stresses and matter flows induced by complex gravitational processes. The evolution of the different patches as a function of cosmic time is rendered via the diverse realizations of the noise term; meanwhile, at any given cosmic time, sampling the ensemble of patches will originate a nontrivial spatial distribution of the various cosmological quantities. Finally, the overall behavior of the Universe will be obtained by averaging over the patch ensemble. We assume a simple and physically reasonable parameterization of the noise term, gauging it against a wealth of cosmological datasets in the local and high-redshift Universe. We find that, with respect to standard $\Lambda$CDM, the ensemble-averaged cosmic dynamics in the $\eta$CDM model is substantially altered by the stochasticity in three main respects: (i) an accelerated expansion is enforced at late cosmic times without the need for any additional exotic component (e.g., dark energy); (ii) the spatial curvature can stay small even in a low-density Universe constituted solely by matter and radiation; (iii) matter can acquire an effective negative pressure at late times. The $\eta$CDM model is Hubble-tension free, meaning that the estimates of the Hubble constant from early and late-time measurements do not show marked disagreement as in $\Lambda$CDM. We also provide specific predictions for the variance of the cosmological quantities among the different patches of the Universe at late cosmic times. Finally, the fate of the Universe in the $\eta$CDM model is investigated, to show that the cosmic coincidence problem is relieved without invoking the anthropic principle.
\end{abstract}


\section{Introduction}\label{sec|intro}

The standard $\Lambda$CDM model of the Universe has proven to be extremely successful in reproducing to a high degree of accuracy many cosmological observations, most noticeably the cosmic microwave background temperature and polarization spectra (e.g., Bennett et al. 2003; Planck collaboration et al. 2013, 2020a), supernovae (SN) I$a$ cosmography (e.g., Perlmutter et al. 1999; Scolnic et al. 2018; Brout et al. 2022), baryon acoustic oscillations (BAO) measurements (e.g., Eisenstein et al. 2005; Beutler et al. 2011; Zhao et al. 2022), cosmic shear galaxy surveys (e.g., Heymans et al. 2013; Amon et al. 2022; Secco et al. 2022), galaxy clusters (e.g., White et al. 1993; Allen et al. 2011; Mantz et al. 2022), and many others (e.g., see recent review by Turner 2022 and references therein; also the essay by Efstathiou 2023). Despite these astonishing successes, the $\Lambda$CDM model maintains a fundamentally empirical character, in that it postulates the existence of a mysterious dark energy component with exotic negative pressure, that at late cosmic times dominates the energy budget and enforces an accelerated expansion of the Universe.

From an observational perspective, the evidence for dark energy remains mainly related to the interpretation of two occurrences: the accelerated expansion of the Universe at late cosmic times, as mainly indicated by type I$a$ SN determinations of the magnitude-redshift diagram; the nearly zero curvature (flat geometry) of a Universe with a low matter (baryons and dark matter) content, as mainly indicated by CMB and BAO data. From a theoretical perspective, the situation is even more dramatic: the value of the present dark energy density required to explain the aforementioned observations is far below the Planck or any natural scale in particle physics; nonetheless, it is of the same order of magnitude with respect to the matter density rather than extremely smaller or fatally larger, thus allowing its observability in this very precise moment of cosmic history (e.g., Zel'dovich 1968; Weinberg 1989). Furthermore, in recent years some discrepancies from the $\Lambda$CDM paradigm started to emerge with various degrees of significance, among which the (in)famous Hubble tension that concerns the disagreement between late-time measurements of the Hubble constant with respect to the $\Lambda$CDM predictions from the CMB (e.g., Efstathiou 2020; Riess et al. 2022; also Di Valentino et al. 2021), and the $S_8$ tension that concerns a deficit in the weak lensing amplitude measured by cosmic shear galaxy surveys with respect to the CMB expectations (e.g., Asgari et al. 2021; Secco et al. 2022; also Amon \& Efstathiou 2022).

In this complex landscape a number of alternative cosmological models have been designed with the aim to interpret the cosmic acceleration without invoking a dark energy component. For example, one may consider modified gravity theories that introduce additional degrees of freedom in the gravitational and/or matter action (see reviews by Clifton et al. 2012; Najiri et al. 2017; Saridakis et al. 2021; also Planck Collaboration et al. 2016), or one may phenomenologically modify the Friedmann equation by additional terms that depend on the matter density in a non-linear way like in the Cardassian scenarios (see Freese \& Lewis 2002; Xu 2012; Magana et al. 2018), or one may alter the mass-energy evolution equation with bulk viscosity terms based on thermodynamical considerations (see Lima et al. 1988; Brevik et al. 2011; Herrera-Zamorano et al. 2020).
Alternatively, the interpretation of SN data may be biased if the observer is located in a local underdense region (e.g., Celerier 2000) or considering that the SN sources tend to be associated with overdensities (e.g., Deledicque 2022).

\begin{figure}[!t]
\centering
\includegraphics[width=1.\textwidth]{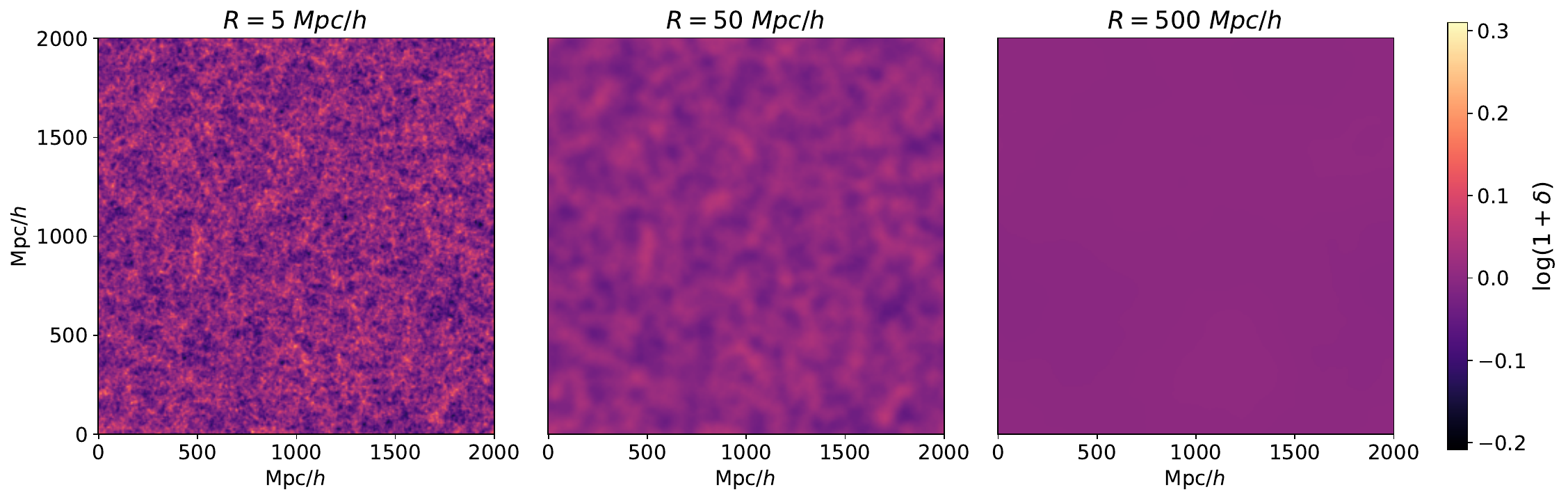}
\includegraphics[width=.95\textwidth]{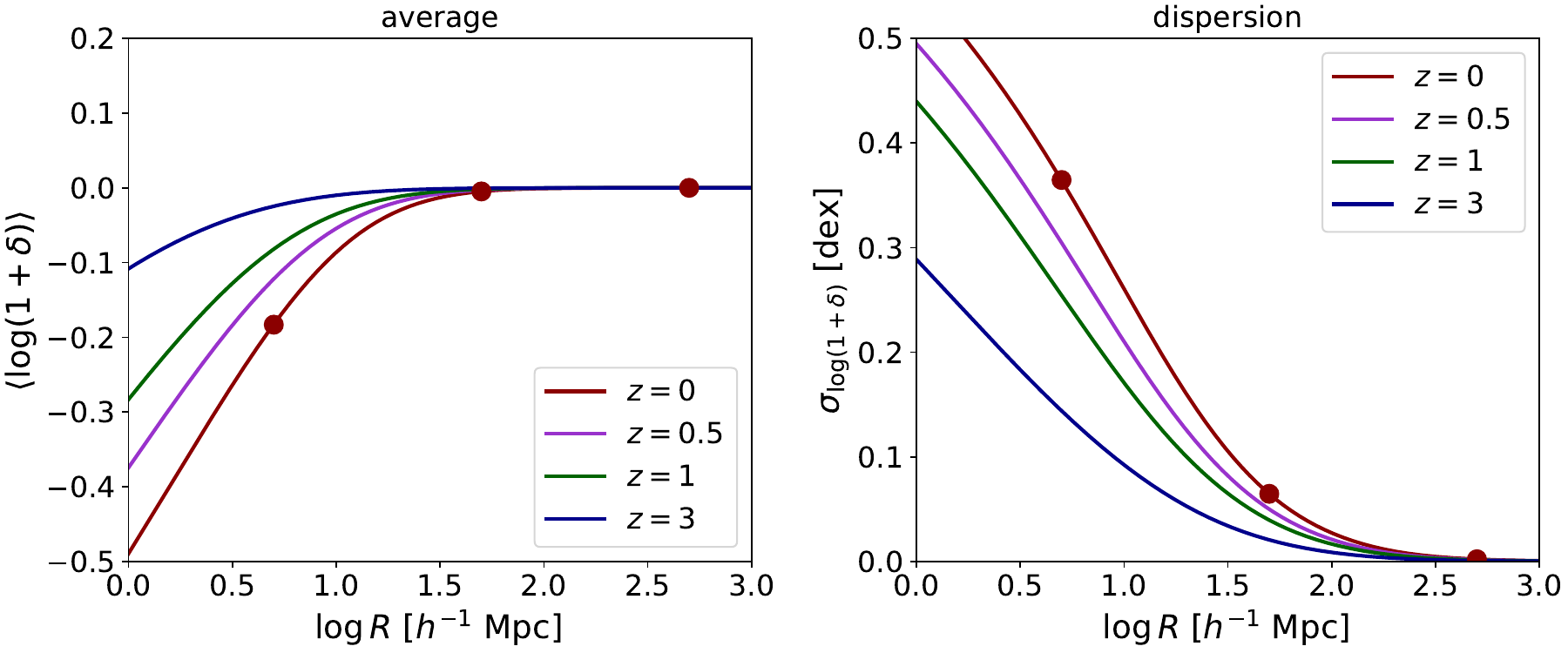}
\caption{Top panels: snapshots at $z\approx 0$ extracted from a $N-$body simulation in the $\Lambda$CDM cosmology with a box of $2\,h^{-1}$ Gpc and $1024^3$ particles, which shows the amplitude $\log (1+\delta)$ of the overdensity field (color-coded) smoothed via a top-hat window function on scales $R\approx 5\, h^{-1}$ Mpc (left), $50\, h^{-1}$ Mpc (middle), $500\, h^{-1}$ Mpc (right). Bottom panel:
average and dispersion of the log-normal distribution followed by the density contrast field in the simulation, when smoothed on different scales $R$, at redshifts $z\approx 0$ (red), $0.5$ (magenta), $1$ (green) and $3$ (blue). The red circles highlight the smoothing scales used in the snapshots shown in the top panels.}\label{fig|Nbody}
\end{figure}

Yet another possibility, more connected to this work, focuses on the role of matter inhomogeneities and anisotropies that may affect the cosmic expansion due to backreaction or statistical sampling effects (e.g., Buchert \& Ehlers 1997; Barausse et al. 2005; Wiltshire 2007; Buchert 2008; Kolb 2011; Buchert \& Rasanen 2012; Clifton 2013; Racz et al. 2017; Cosmai et al. 2019; Schander \& Thiemann 2021). In fact, on the spatial scales associated to cosmic structures, matter and radiation are expected to be affected by a variety of complex physical processes, including local inhomogeneities, clumping, fractality, anisotropic stresses, gravitational and electromagnetic interactions, tidal torques, inflows and outflows, etc. When investigating background cosmology one usually does not care with the fine details of such a behavior, but instead focuses on the description of quantities spatially-averaged over sufficiently large scales, where an homogeneous/isotropic Universe is assumed to hold according to the celebrated cosmological principle. But given the non-linear nature of Einstein's general relativity equations, it is far from clear to what extent the complex gravitational dynamics on smaller scales can be neglected in modeling the evolution of larger-scale patches and of the Universe as a whole.

From an observational perspective, testing homogeneity and isotropy on large scales along cosmic history is not trivial. On the one hand, observations of the CMB show a remarkable consistency with statistical isotropy (meaning that small CMB fluctuations originated from inflation feature the same power spectrum in different directions on the sky), apart from a few `anomalies' on the largest scales (see Planck collaboration 2020b; also Chiocchetta et al. 2021). On the other hand, it may well be that deviations from homogeneity and isotropy can emerge at later cosmic time, thus requiring them to be checked with different probes at lower redshift. In this vein, many studies have been conducted exploiting galaxies (see Javanmardi \& Kroupa 2017; Sarkar et al. 2019), radio-sources (see Bengaly et al. 2019; Siewert et al. 2021; Secrest et al. 2022), gamma-ray bursts (see Tarnopolski 2017; Ripa \& Shafieloo 2019; Andrade et al. 2019; Horvath et al. 2020), AGN/quasars (see Secrest et al. 2021; Goncalves et al. 2021; Friday et al. 2022; Lopez et al. 2022; Tiwari et al. 2023), large-scale bulk flows (see Kashlinsky et al. 2011; Atrio-Barandela et al. 2015; Watkins et al. 2023), HI gas (see Hazra \& Shafieloo 2015; Avila et al. 2018, 2023), X-ray emission from galaxy clusters (see Migkas et al. 2020, 2021), and type I$a$ SN (see Javanmardi et al. 2015; Bernal et al. 2017; Colin et al. 2019; Hu et al. 2020; Krishnan et al. 2022; Rahman et al. 2022; Zhai \& Percival 2022; McConville \& Colgain 2023); the results are still somewhat controversial, with some of these analyses supporting the cosmological principle, and some others claiming statistically significant deviations from it (see review by Kumar Aluri et al. 2023).

One of the most prominent manifestation of the anisotropic and inhomogeneous nature of the gravitationally-driven evolution of the Universe is the progressive appearance of the cosmic web toward late cosmic times. This spider-network of quasi-linear structures that permeates the cosmos on large scales is constituted by anisotropic filamentary and planar features (filaments and sheets), intersecting into nodes (knots) where galaxy clusters tend to reside, and surrounding large underdense regions (voids) which occupy most of the volume. As shown by numerical simulations, the formation and evolution of the cosmic web is driven by gravitational tidal forces induced by inhomogeneities in the mass distribution, and by the multistream migration of matter from adjacent structures (e.g., Springel et al. 2006; Shandarin et a. 2012; Vogelsberger et al. 2014; Libeskind et al. 2018; Martizzi et al. 2019; Wilding et al. 2021; see also Angulo \& Hahn 2022).

It is instructive to highlight a few statistical properties of the cosmic web by running a $N-$body numerical simulation in the standard $\Lambda$CDM cosmology, and in particular by extracting the distribution of the (over)density field $1+\delta\equiv \delta\rho/\rho$ smoothed on different coarse-graining scales $R$. Snapshots of such a simulation at redshift $z\approx 0$ and smoothing scales $R\approx 5-50-500\, h^{-1}$ Mpc are illustrated in the top panels of Fig. \ref{fig|Nbody}. Anisotropic/inhomogeneous conditions are manifest for small $R\lesssim$ a few Mpc and are progressively washed out for larger $R$, to the point of becoming negligible at scales $R\gtrsim$ hundreds Mpc, i.e. an appreciable fraction of the current Hubble radius. Interestingly, simulations show that the emergence of the cosmic web via the gravitationally-driven growth of  structures and voids transforms the statistical distribution of the overdensity field $\Omega\sim 1+\delta$ from the initial Gaussian to a log-normal shape (even on quasi-linear scales), as also pointed out by theoretical and numerical studies in the literature (e.g., Coles \& Jones 1991; Neyrinck et al. 2009; Repp \& Szapudi 2017, 2018). Specifically, the average and dispersion of such lognormal distribution is plotted in the bottom panels of Fig. \ref{fig|Nbody} as a function of the smoothing scale and redshift. Moving from larger to smaller scales, the average $\langle\log(1+\delta)\rangle$ gets progressively biased toward negative values (corresponding to underdense regions, e.g. voids) and the dispersion $\sigma_{\log(1+\delta)}$ increases; the effect is more pronounced toward lower redshifts as structure formation proceeds. From these outcomes it should be clear that scales smaller than a few Mpcs are strongly inhomogeneous and anisotropic, but are dominated by non-linear structures and peculiar motions detached from the Hubble flow, hence any major link with the overall cosmic dynamics is presumably lost. At the other end, scales larger than hundreds Mpcs are fully homogeneous and isotropic, so that averaging cosmological quantities over them brings about the standard dynamics, but at the price of ignoring any possible backreaction effects due to structure formation
from the quasi-linear scales of the cosmic web that are still connected to Hubble flow.

Motivated from the above, in this paper we focus on scales of tens Mpc associated to the cosmic web,
and investigate the impact on the cosmic dynamics of the small residual anisotropies/inhomogeneities
present there. Our framework is inspired by some of the backreation models mentioned above, though we attack the problem via a different approach based on stochastic differential equations. This envisages that such patches of the Universe will undergo slightly different evolutions due to local inhomogeneities, matter flows related to anisotropic stresses, tidal forces, gravitational torques, etc. However, we do not aim to follow the details of such a complex dynamics (that would be practically impossible to handle in semi-analytic terms), and revert instead to an effective statistical description for the evolution of the various patches in terms of the different realization of an appropriate noise term in the mass-energy evolution equation. At any given cosmic time, sampling the ensemble of patches will originate a nontrivial spatial distribution of the various cosmological quantities, while the overall behavior of the Universe will then be obtained by averaging over the patch ensemble.
We assume a simple phenomenological and physically reasonable parameterization of the noise term, tuning it against a wealth of cosmological datasets in the local and high-redshift Universe.
We find that, with respect to $\Lambda$CDM, in our stochastic model an accelerated expansion can be enforced at late cosmic times and the curvature can stay small even in a low-density Universe constituted by matter and radiation; this behavior will turn out to be ultimately due to the nature of the large-scale structure formation, which places far more volume in underdense than in overdense regions, causing the average expansion to skew toward an accelerated behavior. Remarkably, matter can also acquire an effective negative pressure at late times. We also provide  predictions for the variance
of the cosmological quantities among the different patches as induced by the noise. Finally, we show that our model is capable of solving the Hubble tension and relieving the coincidence problem.

The plan of the paper is as follows. In Sect. \ref{sec|basics} we introduce our cosmological model; in Sect. \ref{sec|deterministic} we investigate its implication on the cosmic dynamics; in Sect. \ref{sec|cosmofit} we gauge the noise term against a wealth of cosmological datasets; in Sect. \ref{sec|stochasticity} we examine the time-dependent variance implied by the noise on the cosmological quantities; in Sect. \ref{sec|destiny} we explore the future evolution of the Universe; in Sect. \ref{sec|discussion} we provide answers to some frequently asked questions; finally, in Sect. \ref{sec|summary} we summarize our findings and outline future perspectives.

\section{Basic setup}\label{sec|basics}

We consider a Universe composed by (baryonic$+$dark) matter and radiation, subject to the laws of canonical general relativity. We focus on patches of order tens Mpc, where the local density field is unbiased with respect to the average density of the Universe (i.e., $\langle\log (1+\delta)\rangle\approx 0$), but residual small anisotropies/inhomogeneities, associated to the cosmic web, are present (see Sect. \ref{sec|intro} and Fig. \ref{fig|Nbody}); in these conditions we assume that the standard Friedmann-Robertson-Walker metric can be applied to a good approximation (as also shown by detailed GR simulations; see Rasanen 2010; Koksbang 2019; Macpherson et al. 2019; Adamek et al. 2019). Yet any of such patches of the Universe will experience slightly different evolutions along the cosmic history due to local inhomogeneities, matter flows induced by anisotropic stresses, tidal forces, and other complex gravitational processes that for all practical purposes are extremely difficult to model ab-initio or to handle (semi-)analytically. Therefore we revert to a statistical description in terms of stochastic differential equations, phenomenologically characterizing the different evolution of the patches as a function of cosmic time via the diverse realizations of a noise term; meanwhile, at any given cosmic time, sampling the ensemble of patches will originate a nontrivial spatial distribution of the various cosmological quantities. Finally, the overall behavior of the Universe at any comic time will be obtained by averaging over the patch ensemble.

Technically, we add a simple stochastic term to the mass-energy evolution equation, with the following properties (see also Sect. \ref{sec|discussion} for an extended discussion): stochasticity is driven by a Gaussian white noise $\eta(t)$; the stochastic term scales proportionally to the energy density of each component, so that linearity of the mass-energy evolution equation is preserved; the stochastic term features an inverse power-law dependence on the Hubble parameter, so that at early times random effects become negligible, in order to be consistent with the statistical isotropy of the CMB. The evolution of each different patch of the Universe is described by the diverse realizations of the noise term $\eta(t)$ via the system of stochastic differential equations (in the Stratonovich sense, see Appendix \ref{sec|app_stocha}):
\begin{equation}\label{eq|friedrand}
\left\{
\begin{aligned}
H^2 & = \cfrac{8\pi G}{3}\, (\rho_m+\rho_\gamma)-\cfrac{\kappa\,c^2}{a^2}~,\\
\\
\dot\rho_{m,\gamma} & = -3\, H\, \rho_{m,\gamma}\,(1+w_{m,\gamma})+\zeta\, \rho_{m,\gamma}\, \sqrt{H_\star}\, \left(\cfrac{H}{H_\star}\right)^\alpha\, \eta(t)~;
\end{aligned}
\right.
\end{equation}
here $G$ is the standard gravitational constant and $c$ the speed of light, $a(t)$ is the scale factor normalized to unity at present time, $H(t)\equiv \dot a/a$ is the Hubble rate and $H_\star\equiv 100$ km s$^{-1}$ Mpc$^{-1}$ a reference value, $\kappa$ is the curvature constant, and $\rho_{m,\gamma}$ the energy density of matter and radiation with equation of state parameters $(w_m;w_\gamma)=(0;1/3)$. In addition, as mentioned above $\eta(t)$ is a Gaussian white noise (physical dimensions $1/\sqrt{t}$) with ensemble-average properties $\langle \eta(t)\rangle=0$ and  $\langle\eta(t)\eta(t')\rangle=2\, \delta_D(t-t')$, and ($\zeta\geq 0$; $\alpha\leq 0$) are two parameters regulating the strength and redshift dependence of the noise, yet to be specified (these will be set by comparison with cosmological observables, see Sect. \ref{sec|cosmofit}). Note that when introducing the normalized Hubble rate $h\equiv H/H_\star$ and the density parameters $\Omega_{m,\gamma}\equiv 8\pi G\,\rho_{m,\gamma}/3\, H^2$, the Friedmann equation just reads $\Omega_\kappa\equiv -\kappa\, c^2/a^2\, h^2 = 1-\Omega_m-\Omega_\gamma$.

The multiplicative factor in front of the Gaussian white noise $\eta$ is an ansatz, inspired by the following naive argument. As it occurs in $\Lambda$CDM (see Fig. \ref{fig|Nbody}), one expects that at late times the overdensity $1+\delta\propto \rho$ smoothed on a certain coarse-graining scale (e.g., tens of Mpcs) constitutes a random field following a lognormal distribution (see also theoretical arguments by Coles \& Jones 1991; also Neyrinck et al. 2009 and Repp \& Szapudi 2017, 2018); in simple words, sampling patches of the present Universe at different spatial locations yields a lognormal distribution for $\rho$ or equivalently a normal distribution for $\log \rho$.
In terms of basic stochastic processes, such a distribution would be naturally originated by an ensemble of regions whose density evolves stochastically in time under a Gaussian white noise $\eta(t)$, or in other words for which ${\rm d}_t\log \rho\sim \dot \rho/\rho\propto \eta(t)$, with the proportionality constant being related to the variance of the density distribution (see Risken 1996, Paul \& Baschnagel 2013; also Appendix A). Although Eqs.~(\ref{eq|friedrand}) describe a more complex stochastic system (since the second equation features a dilution term and it is coupled to the first via the Hubble parameter), this analogy has inspired us to adopt a noise term $\dot \rho\sim \zeta\,\,H^\alpha\, \rho$, with the parameters $\zeta$ and $\alpha$ describing our ignorance on the present value and on the redshift evolution of the variance in the density distribution for a generic cosmology that can be in principle different from $\Lambda$CDM.

We stress that Eqs.~(\ref{eq|friedrand}) should be meant to hold on patches of the Universe with a typical smoothing (or coarse-graining) scale, that constitutes in itself a hidden parameter of the model, though in turn fully determined by the noise and cosmological ones. Such a scale will be estimated after setting the noise and cosmological parameters via comparison with data (Sect. \ref{sec|results}) and evaluating the typical fluctuations of the density field induced by the noise (Sect. \ref{sec|stochasticity}); here we anticipate that it will turn out to be around tens Mpc, a typical size associated to the cosmic web. In the same perspective, notice that the noise term in Eqs.~(\ref{eq|friedrand}) subtends deviations from local energy conservation in different patches of the Universe with size given by the aforementioned coarse-graining scale, that can associated, e.g., to matter flows; given the amount of fluctuations in the density field induced by the noise as computed in Sect. \ref{sec|stochasticity}, these deviations will turn out to be minor.

Finally, note that the noise term has a natural timescale $t_\eta$ associated with it, that can be easily derived by dimensional analysis of the second term on the r.h.s. of the mass-energy evolution equation above; writing $\dot \rho\sim \rho/t_\eta$ and $\eta\sim 1/\sqrt{t_\eta}$ one finds that $t_\eta\sim 1/(H_\star\, \zeta^2\, h^{2\alpha})$. This has to be compared with the typical timescale of deterministic dilution expressed by the first term on the r.h.s. of the same equation, e.g. for matter $t_{\rm exp}\sim 1/(3\,H)$. The competition between noise and dilution can be quantified by the ratio $t_\eta/t_{\rm exp}\sim 3/(\zeta^2\,h^{2\alpha-1})$, that will be evaluated in Sect. \ref{sec|results} after having determined the noise and cosmological parameters via comparison with data.

Hereafter we will refer to this model of the Universe as $\eta$CDM since $\eta$ is the standard mathematical symbol for the noise ruling its dynamics. Although constituting a seemingly simple modification to the standard cosmological framework, our proposal has relevant implications for the cosmic history. Three preliminary remarks are in order. First, since Eqs.~(\ref{eq|friedrand}) above are coupled, not only $\rho_{m,\gamma}$ but also the quantities $a$ and $H$ are promoted to stochastic variables; these will fluctuate in time under the action of the noise, in a slightly different way for each of the patches (corresponding to different noise realizations). This is better highlighted by combining Eqs.~(\ref{eq|friedrand}) to obtain the acceleration equation
\begin{equation}\label{eq|friedacc}
\frac{\ddot{a}}{a} = -\frac{4\pi\, G}{3}\,(\rho_m+2\,\rho_\gamma) + \zeta\, \frac{4\pi\, G}{3\, \sqrt{H_\star}}\, (\rho_m+\rho_\gamma)\, \left(\frac{H}{H_\star}\right)^{\alpha-1}\, \eta(t)~,
\end{equation}
implying that the noise term acts as a random fluctuation onto the dynamics in each patch of the Universe; plainly, posing $\zeta=0$ yields the usual acceleration equation.
Therefore, under the influence of the noise, each patch of the Universe will undergo a slightly different evolution in cosmic time; meanwhile, at any given cosmic time, sampling the ensemble of patches will originate a nontrivial spatial distribution of the various cosmological quantities. Second, note that the noise term is `multiplicative', meaning that it depends on the system state: as $\eta(t)$ fluctuates, also the variables $\rho_{m,\gamma}$ and $H$ appearing in the stochastic term vary, therefore in Eq.~(\ref{eq|friedacc}) one finds that $\langle\rho_{m,\gamma}\,H^{\alpha-1}\,\eta\rangle$ is not null even if $\langle\eta\rangle$ is; the result will be a noise-induced drift affecting the late-time ensemble-averaged cosmological evolution (actually accelerating it, as shown in Sect. \ref{sec|deterministic}).

Third, plainly the overall cosmic dynamics will be specified not only by the above equations, but also by sensible boundary conditions. Since the noise term is by construction negligible at early times, the evolution should mirror that of a standard cosmological model in the remote past. Then one may envisage of integrating the stochastic system forward in cosmic time from such initial conditions; however, this procedure cannot be correct, since it would possibly originate an uncontrolled diffusion of the cosmological quantities toward the present, to values that in principle may be very far from the spatially-average ones or even nonphysical in some intermediate step of the cosmic history. To avoid the issue, one must require that the values of the cosmological quantities measured by an observer here and now are close to the average values; this means that the overall evolution cannot be a simple diffusion, but rather a diffusion bridge, i.e., a controlled diffusion such that the initial and final values of the different random paths are appropriately assigned (we anticipate that the final values will be set by comparison with cosmological data, and the initial one by integrating back in time the equations describing the average evolution; more on this below). Diffusion bridges have recently found many applications ranging from genetics, to economics, to data science (see Pedersen 1995; Durham \& Gallant 2002; Delyon \& Hu 2006; Lindstrom 2012; Bladt \& Sorensen 2014; Whitaker et al. 2017; Heng et al. 2022). The treatment of such conditioned diffusion problems is not trivial at all (especially when drift terms are present), and exact solutions are practically impossible to obtain except for very peculiar setups; however, various techniques have been developed to find approximate solutions in general cases, as exploited below.

Before proceeding, it is convenient to put Eq.~(\ref{eq|friedrand}) in a more transparent and numerically tractable form. First of all, we differentiate the Friedmann equation, re-express the term $\kappa\,c^2/a^2$ in terms of $H$ and $\rho_{m,\gamma}$, exploit the mass-energy evolution equations to write explicitly $\dot\rho_{m,\gamma}$, and then use $\dot\Omega_{m,\gamma}/\Omega_{m,\gamma}=\dot\rho_{m,\gamma}/\rho_{m,\gamma}-2\,\dot h/h$ to obtain an equation for the density parameters $\Omega_{m,\gamma}$. Finally, we introduce the normalized Hubble rate $h\equiv H/H_\star$ and redefine the time variable $\tau\equiv H_\star\, t$ such that $\eta(t)\rightarrow \eta(\tau)\,\sqrt{H_\star}$ by the property of the Gaussian noise. All in all, we get (overdot means differentiation with respect to $\tau$):
\begin{equation}\label{eq|friednormrand}
\left\{
\begin{aligned}
\dot h &= h^2\,\left(-1-\cfrac{\Omega_m}{2}-\Omega_\gamma\right)+\cfrac{\zeta}{2}\,(\Omega_m+\Omega_\gamma)\, h^{\alpha+1}\, \eta(\tau)\\
\\
\dot \Omega_m &=\Omega_m\, h\, (-1+\Omega_m+2\,\Omega_\gamma)+\zeta\,(1-\Omega_m-\Omega_\gamma)\, \Omega_m\, h^\alpha\, \eta(\tau)\\
\\
\dot \Omega_\gamma &=\Omega_\gamma\, h\, (-2+\Omega_m+2\,\Omega_\gamma)+\zeta\,(1-\Omega_m-\Omega_\gamma)\, \Omega_\gamma\, h^\alpha\, \eta(\tau)~,
\end{aligned}
\right.
\end{equation}
supplemented by the final boundary values $(h_0,\Omega_{m,0},\Omega_{\gamma,0})$ at the present time $\tau_0$. Plainly for
$\zeta=0$ the usual dynamics of a (non-zero curvature) patch of the Universe is recovered.
As mentioned above, in principle a stochastic system just characterized by a terminal boundary value is ill-defined, since the randomness naturally develops with the evolution proceeding forward in time. However, in the present context the problem is made meaningful by the temporal behavior of the noise term, which makes fluctuations negligible at early times, to imply that the full stochastic solutions must converge in the remote past to the spatially-average one (i.e., obtained at each cosmic time by averaging over the patch ensemble). Thus the technique to solve the problem (mutuated by Whitaker et al. 2017; details in Appendix \ref{sec|app_bridge}) is to separate the stochastic variables $(h;\Omega_m;\Omega_\gamma)=(\bar h+\tilde h;\bar{\Omega}_{m}+\tilde{\Omega}_{m};\bar{\Omega}_{\gamma}+\tilde{\Omega}_{\gamma})$ in an average (barred variables) and in a residual random (tilded variables) component. On the one hand, the average behavior can be shown to satisfy an ordinary differential equation that is easily solved backward in time from the terminal condition at $\tau_0$, to provide an initial condition at a time $\tau_{\rm in}<<\tau_0$ for the full system; remarkably, this equation inherits a noise-induced drift term that will have relevant consequences on the late-time cosmic dynamics (see next Section). On the other hand, the residual random component will render the fluctuations of the various patches, and will originate variance in the (spatial) distribution of the cosmological quantities at a given cosmic time. Such a random component can be shown to satisfy a stochastic equation (actually also requiring the average solution as an input) that can be integrated forward in time from an initial null value at $\tau_{\rm in}$; moreover, this equation features a spurious drift term that forces the solution to hit a null terminal value at $\tau_0$, i.e. to execute a diffusion bridge (see Appendix \ref{sec|app_bridge} for details). All in all, the overall stochastic process will be characterized by well-defined initial and final values connected by the average evolution plus some random behavior at intermediate times.

In the next Sections we will apply such a technique to Eqs.~(\ref{eq|friednormrand}) to investigate the cosmic dynamics, to set the terminal conditions and the noise parameters by comparison with cosmological observables, and to estimate the variance in the evolution of the different patches as induced by the noise term.

\section{Average evolution}\label{sec|deterministic}

The equations ruling the ensemble-averaged behavior $(\bar h,\bar{\Omega}_{m},\bar{\Omega}_{\gamma})$ in the evolution of Eqs.~(\ref{eq|friednormrand}) can be derived from the procedure outlined in Appendix \ref{sec|app_stocha}. After some tedious algebra we obtain
\begin{equation}\label{eq|friednew}
\left\{
\begin{aligned}
\dot{\bar{h}} &= \bar{h}^2\,\left(-1-\cfrac{\bar{\Omega}_m}{2}-\bar{\Omega}_\gamma\right)+\cfrac{\zeta^2}{2}\,(\bar{\Omega}_m+\bar{\Omega}_\gamma)\, \bar{h}^{2\alpha+1}\, \left[1-\cfrac{1-\alpha}{2}\, (\bar{\Omega}_m+\bar{\Omega}_\gamma)\right]\\
\\
\dot{\bar{\Omega}}_m &=\bar{\Omega}_m\, \bar{h}\, (-1+\bar{\Omega}_m+2\,\bar{\Omega}_\gamma)+\zeta^2\,(1-\bar{\Omega}_m-\bar{\Omega}_\gamma)\, \bar{\Omega}_m\, \bar{h}^{2\alpha}\, \left[1-\cfrac{4-\alpha}{2}\, (\bar{\Omega}_m+\bar{\Omega}_\gamma)\right]\\
\\
\dot{\bar{\Omega}}_\gamma &=\bar{\Omega}_\gamma\, \bar{h}\, (-2+\bar{\Omega}_m+2\,\bar{\Omega}_\gamma)+\zeta^2\,(1-\bar{\Omega}_m-\bar{\Omega}_\gamma)\, \bar{\Omega}_\gamma\, \bar{h}^{2\alpha}\, \left[1-\cfrac{4-\alpha}{2}\, (\bar{\Omega}_m+\bar{\Omega}_\gamma)\right]~,
\end{aligned}
\right.
\end{equation}
where on the right hand side the second addenda are the noise-induced drift terms, stemming from the multiplicative nature of the noise; in fact, the average evolution is informed on, and affected by the noise of the original stochastic process. This system of equations may be easily solved backward in cosmic time from terminal conditions at the present epoch $\tau_0$ given by $(h_0;\Omega_{m,0};\Omega_{\gamma,0})$, to yield the average evolution of the Hubble rate and density parameters across cosmic history. We are now ready to point out that the dynamics implied by Eqs.~(\ref{eq|friednew}) has three relevant consequences.

\begin{itemize}

\item \emph{Late-time acceleration without dark energy.} One can derive the deceleration parameter $q$ from the first of Eqs. (\ref{eq|friednew}), to read $\dot{\bar{h}}/\bar{h}^2\equiv -(1+q)$. At late times one can neglect radiation ($\bar{\Omega}_\gamma<<\bar{\Omega}_m$) so that
\begin{equation}\label{eq|q}
\bar{q} = -1-\cfrac{\dot{\bar{h}}}{\bar{h}^2} = \cfrac{\bar{\Omega}_m}{2}-\cfrac{\zeta^2}{2}\,\bar{\Omega}_m\,\bar{h}^{2\alpha-1}\,\left(1-\cfrac{1-\alpha}{2}\, \bar{\Omega}_m\right)~;
\end{equation}
whenever $\bar{\Omega}_m\lesssim 2/(1-\alpha)$ the stochastic term tends to reduce the deceleration parameter expected from a standard open cosmology without dark energy, and for appropriate value of $\alpha$ and $\zeta$ an accelerated expansion with $q<0$ at late times can be enforced. The physical interpretation of this effect is that the overall ensemble of patches tends to drift toward an evolution dominated by low-density regions (see also Sect. \ref{sec|discussion}). High-order cosmographic quantities can be easily computed from the expression above (e.g., Visser 2005); e.g., the jerk parameter is given by
\begin{equation}\label{eq|j}
\bar{j}=1+3\,\cfrac{\dot{\bar{h}}}{\bar{h}^2}+\cfrac{\ddot{\bar{h}}}{\bar{h}\,\bar{h}^3} = \bar{q}+2\, \bar{q}^2-\cfrac{\dot{\bar{q}}}{\bar{h}}~,
\end{equation}
and will be shown to have a non-trivial evolution at late cosmic times.

\item \emph{Small curvature in a low-density Universe.} One can derive the average evolution of the global effective curvature $\bar{\Omega}_\kappa$ along the following lines. By combining Eqs.~(\ref{eq|friednew}) after some algebraic manipulation one gets
\begin{equation}\label{eq|kappa_aux}
\cfrac{2\,\dot{\bar{h}}}{\bar{h}} = \cfrac{\dot{\bar{\Omega}}_m+\dot{\bar{\Omega}}_\gamma}{1-\bar{\Omega}_m-\bar{\Omega}_\gamma}-2\, \bar{h}+\,\cfrac{3}{2}\,\zeta^2\,\bar{h}^{2\alpha} (\bar{\Omega}_m+\bar{\Omega}_\gamma)^2~.
\end{equation}
Formally integrating in time (recall that $\bar{h}\,{\rm d}t={\rm d}\ln \bar{a}$ in terms of the scale factor) and recognizing that the integration constant is related to the curvature parameter $\bar{\Omega}_\kappa\propto -\kappa/\bar{h}^2\,\bar{a}^2$ one obtains the modified Friedmann constraint
\begin{equation}\label{eq|kappa}
\bar{\Omega}_\kappa = (1-\bar{\Omega}_m-\bar{\Omega}_\gamma)\, \exp\left[-\cfrac{3}{2}\,\zeta^2\,\int{\rm d}\ln \bar{a}~\bar{h}^{2\alpha-1}\,(\bar{\Omega}_m+\bar{\Omega}_\gamma)^2\right]~.
\end{equation}
Eq.~(\ref{eq|kappa}) directly implies that $\bar{\Omega}_\kappa$ is negligible in the remote past, since the exponential term containing the noise parameters tends to one and $\bar{\Omega}_m+\bar{\Omega}_\gamma\approx 1$ applies like in the standard $\Lambda$CDM cosmology. However, $\bar{\Omega}_k$ starts growing toward the present, then attains a maximum value $\bar{\Omega}_\kappa\lesssim 0.1$ at the time when $\bar{h} \approx [\zeta^2\,(1-(1-\alpha)\,\bar{\Omega}_m/2)]^{1/(1-2\alpha)}$ and eventually decreases again toward zero in the infinite future. On the one hand, such an evolution is consistent with inflationary scenarios, which predict that any possible initial curvature has been erased via a super-exponential expansion already at a time around $10^{-32}$ s after the Big Bang (see Efstathiou \& Gratton 2020); on the other hand, the noise induces via Eq.~(\ref{eq|kappa}) a non-trivial evolution of the curvature parameter implying modest deviations from flatness around the cosmic times where the acceleration sets in, that could possibly constitute a specific test of the $\eta$CDM model in the future. All in all, for appropriate value of $\zeta$ and $\alpha$ the curvature parameter $\bar{\Omega}_{\kappa,0}$ at present may be appreciably reduced to small values, even in a low-density matter-dominated Universe.

\item \emph{Matter with negative pressure at late times.} One can derive the effective equation of state for the matter component at late times from the first two of Eqs.~(\ref{eq|friednew}); neglecting radiation and coming back to the volume energy density via $\dot{\bar{\rho}}_{m}/\bar{\rho}_{m}=\dot{\bar{\Omega}}_{m}/\bar{\Omega}_{m}+2\,\dot{\bar{h}}/\bar{h}$ one has
\begin{equation}\label{eq|wm_aux}
\dot{\bar{\rho}}_m = -3\, \bar{h}\, \bar{\rho}_m + \cfrac{\zeta^2}{2}\,\bar{h}^{2\alpha}\, \bar{\rho}_m\,[2-(4-\alpha)\, \Omega_m+3\, \Omega_m^2]~.
\end{equation}
Then introducing an effective equation of state parameter $\bar{w}_m$ such that $\dot{\bar{\rho}}_m = -3\, \bar{h}\, \bar{\rho}_m\, (1+\bar{w}_m)$, one can write
\begin{equation}\label{eq|wm}
\bar{w}_m= -\cfrac{\zeta^2}{6}\,\bar{h}^{2\alpha-1}\, [2-(4-\alpha)\, \bar{\Omega}_m+3\, \bar{\Omega}_m^2]~;
\end{equation}
this plainly tends to the standard $\bar{w}_m\approx 0$ at early times, while values $\bar{w}_{m,0}<0$ can apply toward the present, implying that matter effectively
behaves as a negative pressure component. However, note from Eq.~(\ref{eq|q}) that in $\eta$CDM the condition $\bar{w}_m<-1/3$ (e.g., violation of the strong-energy condition in general relativity) is not required to enforce cosmic acceleration.

\end{itemize}

\section{Tuning the noise}\label{sec|cosmofit}

Now we move to determine the terminal conditions and the parameters $(\zeta;\alpha)$ regulating the noise strength and redshift dependencies by comparing the evolution implied by Eqs.~(\ref{eq|friednew}) with data. Since the early evolution of the Universe in the $\eta$CDM and $\Lambda$CDM models are indistinguishable (noise is negligible at early times), for the sake of simplicity we set the radiation energy density parameter $\Omega_{\gamma,0}\,h_0^2\approx 2.47\times 10^{-5}$ to the value measured by the \textit{Planck} mission (Planck collaboration et al. 2020a) and the baryon density $\Omega_{b,0}\, h_0^2\approx 0.0222$ to the value suggested by Big Bang Nucleosynthesis constraints (Aver et al. 2015). Then our cosmological model will be characterised by the parameter sets $(h_0;\Omega_{m,0};\zeta;\alpha)$, that we determine by fitting a number of observables in the local and distant Universe; specifically, we consider the following datasets.

\begin{itemize}

\item \emph{Type I$a$ SN with Cepheid zero-point calibration}. We exploit the Pantheon$+$ sample of $\approx 1700$ type I$a$ SN in the redshift range $z\sim 0.001-2.3$ with Cepheid zero-point calibration from the SH$0$ES team (Scolnic et al. 2022; Brout et al. 2022; Riess et al. 2022) to fit for the distance modulus $\mu(z) = 5\, \log(D_L/\mathrm{Mpc})+25$, where the luminosity distance in a positively-curved Universe is computed as
    \begin{equation}
    D_L(z)=\cfrac{c\,(1+z)}{H_0\,\sqrt{\Omega_{\kappa,0}}}\, \sinh\left[\sqrt{\Omega_{\kappa,0}}\,\int_0^z{\rm d}z'\,  \cfrac{H_0}{H(z')}\right]~.
    \end{equation}
    The full covariance matrix of the Pantheon$+$ data has been exploited (this includes statistical and systematic uncertainties in the distance modulus, and Cepheid host covariance).

\item \emph{CMB first-peak angular scale}. We require the model to reproduce the angular scale of the first peak in the CMB temperature spectrum $\theta_\star\approx r_\star/D_M(z_\star)$ as measured by the \emph{Planck} collaboration (2020a); here $r_\star$ is the comoving sound horizon at recombination and $D_M(z_\star)$ is the transverse comoving distance at the recombination redshift $z_\star$. We use the approximations $z_\star(h_0,\Omega_{m,0},\Omega_{b,0})$ and $r_\star(h_0,\Omega_{m,0},\Omega_{b,0})$ by Hu \& Sugiyama (1996).

\item \emph{Baryon acoustic oscillations}. We exploit $18$ datapoints in the redhsift range $z\sim 0.1-2.4$ from various BAO isotropic measurements (Beutler et al. 2011; Kazin et al. 2014; Ross et al. 2015; Alam et al. 2017; Ata et al. 2018; du Mas de Bourboux et al. 2020; Hou et al. 2021; Raichoor et al. 2021; de Mattia et al. 2021; Bautista et al. 2021; Zhao et al. 2022) and fit for the ratio $r_d/D_V(z)$ between the sound horizon at the drag epoch $z_d$ and the angle-averaged galaxy BAO measurement $D_V(z)=\left[c\, z\, D_M^2(z)/H(z)\right]^{1/3}$, where $D_M(z)=D_L(z)/(1+z)$ is the transverse comoving distance. We also consider the ratio between the Hubble distance and the sound horizon at the drag epoch $c/H(z)\,r_d$ inferred from transverse BAO$_\perp$ measurements (see Alam et al. 2017; du Mas de Bourboux et al. 2020; Hou et al. 2021; Bautista et al. 2021; de Mattia et al. 2021), taking properly into account the covariance with some of the isotropic measurements mentioned above. We use the approximations $z_d(h_0,\Omega_{m,0},\Omega_{b,0})$ and $r_d(h_0,\Omega_{m,0},\Omega_{b,0})$ by Aubourg et al. (2015; see also Eisenstein \& Hu 1998).

\item \emph{Cosmic chronometers (CC)}. We also consider the redshift-dependent Hubble parameter $H(z)$ as determined from differential ages of early-type galaxies; the dataset includes $33$ datapoints in the redshift range $z\sim 0.07-2.4$ from various authors (see Simon et al. 2005; Stern et al. 2010; Zhang et al. 2014; Moresco et al. 2012a, 2012b, 2016; Moresco 2015; Ratsimbazafy et al. 2017; Borghi et al. 2022; Jiao et al. 2023). We use the full covariance matrix taking into account modeling uncertainties, mainly related to the choice of the initial mass function, of stellar libraries and stellar population synthesis models (see Moresco et al. 2022 for details). Notice that this datasets is characterized by considerable systematic and statistical uncertainties, and as such will not crucially impact the determination of cosmological parameters.

\item \emph{Age of globular clusters}. We include the latest estimate on the age of the Universe from globular cluster dating (Valcin et al. 2021); however, since the latter has still large statistical uncertainties (and may be affected by several systematics), we also include a hard lower bound $\gtrsim 11$ Gyr to the age of the Universe from classic globular cluster age constraints (e.g., Krauss \& Chaboyer 2004).

\end{itemize}

For parameter inference, we exploit a Bayesian MCMC framework, numerically implemented via the Python package \texttt{emcee} (Foreman-Mackey et al. 2013). We use a standard Gaussian likelihood $\mathcal{L}(\theta)\equiv -\sum_i \chi_i^2(\theta)/2$
where $\theta=\{h_0;\Omega_{m,0};\zeta;\alpha\}$ is the vector of parameters, and the summation is over different observables; for the latter, the corresponding $\chi_i^2= \sum_j [\mathcal{M}(z_j,\theta)-\mathcal{D}(z_j)]^2/\sigma_{\mathcal{D}}^2(z_j)$ is obtained by comparing our empirical model expectations $\mathcal{M}(z_j,\theta)$ to the data $\mathcal{D}(z_j)$ with their uncertainties $\sigma_{\mathcal{D}}^2(z_j)$, summing over the different redshifts $z_j$ of the datapoints (when necessary we take into account the full covariance matrix of the observables). We adopt flat priors $\pi(\theta)$ on the parameters within the ranges $h_0\in [0,1]$, $\Omega_{m,0}\in [0,1]$, $\zeta\in [0,3]$, and $\alpha \in [-3,1]$.  We then sample the posterior distribution $\mathcal{P}(\theta)\propto \mathcal{L}(\theta)\, \pi(\theta)$ by running $\texttt{emcee}$ with $10^4$ steps and $100$ walkers; each walker is initialized with a random position uniformly sampled from the (flat) priors. To speed up convergence, we adopt a mixture of differential evolution (see Nelson et al. 2014) and snooker (see ter Braak \& Vrugt 2008) moves of the walkers, in proportion of $0.8$ and $0.2$, respectively.
After checking the auto-correlation time, we remove the first $20\%$ of the flattened chain to ensure the burn-in; the typical acceptance fractions of the various runs are around $30-40\%$.

\subsection{Fitting Results}\label{sec|results}

\begin{figure}[!t]
\centering
\includegraphics[width=0.85\textwidth]{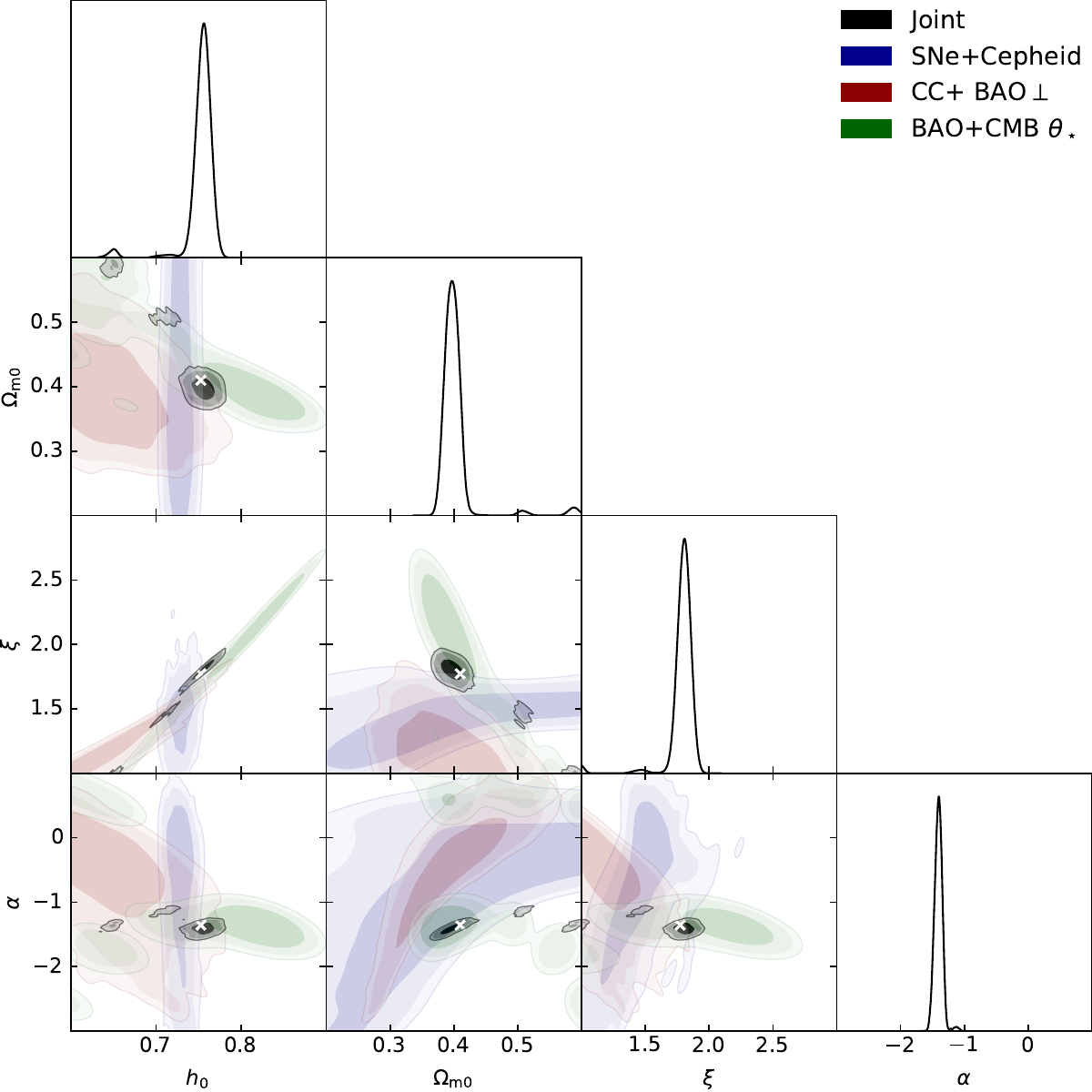}
\caption{MCMC posterior distributions in the $\eta$CDM model, for the normalized Hubble constant $h_0$,
for the matter density parameter $\Omega_{m,0}$, and the parameters regulating the noise strength $\zeta$ and time-dependence $\alpha$. Colored contours/lines refer to different observables: blue for SN + Cepheid, orange for CC + transverse BAO$_\perp$, green for BAO + CMB first peak angular position, and black for joint analysis. The contours show $1-2-3\sigma$ confidence intervals, crosses mark the maximum likelihood estimates of the joint analysis, and the marginalized distributions of the joint analysis are reported on the diagonal panels in arbitrary units (normalized to 1 at their maximum value).}\label{fig|MCMC}
\end{figure}

The outcomes of the fitting procedure are illustrated in the cornerplot of Fig. \ref{fig|MCMC}. The colored contours are the $1-2-3\sigma$ confidence intervals of the posterior for the analysis based on the different datasets: SN+Cepheids (blue), CC+ transverse BAO$_\perp$ (red), BAO + CMB $\theta_\star$ (green), and joint analysis (black); the white crosses are the bestfit position of the joint analysis, and on the diagonal panels the marginalized distributions of the various parameters for the joint analysis are also shown. The marginalized constraints are summarized in Table \ref{Table|MCMC}, where the reduced $\chi^2_r$ of the fits are also reported.

\begin{deluxetable*}{lccccccccccccccccccccccccc}\label{Table|MCMC}
\tablecaption{Marginalized posterior estimates in terms of mean and $1\sigma$ confidence interval [and bestfit value] for the fits with the $\eta$CDM model to different cosmological datasets, as listed in the first column. Other columns report the values of the normalized Hubble constant $h_0$, present matter energy density $\Omega_{m,0}$, noise strength $\zeta$ and noise redshift dependence $\alpha$, and the reduced $\chi_r^2$ of the various fits.}
\tablewidth{0pt}
\tablehead{\colhead{Dataset} & & \colhead{$h_0$} & \colhead{$\Omega_{m,0}$} & \colhead{$\zeta$} & \colhead{$\alpha$} & \colhead{$\chi^2_r$}}
\startdata
\\
\\
Joint & &$0.752^{+0.012}_{-0.005}$ [0.75] & $0.403^{+0.005}_{-0.018}$ [0.40] & $1.78^{+0.08}_{-0.03}$ [1.77] & $-1.41^{+0.07}_{-0.07}$ [-1.36] & 0.44\\
\\
SNe+Cepheid & &$0.73^{+0.01}_{-0.01}$ [0.73] & $0.47^{+0.14}_{-0.27}$ [0.40] & $1.45^{+0.20}_{-0.13}$ [1.44] & $-0.74^{+1.10}_{-0.48}$ [-0.96] & 0.29\\
\\
CC+BAO$_\perp$ & &$0.64^{+0.05}_{-0.05}$ [0.64] & $0.39^{+0.06}_{-0.06}$ [0.39] & $1.14^{+0.20}_{-0.20}$ [1.20] & $-0.48^{+0.71}_{-0.44}$ [-0.41] & 0.46\\
\\
BAO+CMB & &$0.76^{+0.11}_{-0.22}$ [0.84] & $0.42^{+0.06}_{-0.06}$ [0.38] & $1.69^{+0.50}_{-1.69}$ [2.31] & $-1.13^{+0.50}_{-0.50}$ [-1.38] & 1.57\\
\\
\\
\enddata
\tablecomments{We set the radiation density $\Omega_{\gamma,0}\,h_0^2\approx 2.47\times 10^{-5}$ to the value measured by the \textit{Planck} mission (Planck collaboration et al. 2020a) and the baryon density $\Omega_{b,0}\, h_0^2\approx 0.0222$ to the value suggested by Big Bang Nucleosynthesis constraints (Aver et al. 2015); a hard bound on the age of the Universe at $11$ Gyr was placed.}
\end{deluxetable*}

The joint analysis robustly constrains all the parameters of the $\eta$CDM model. As expected the value of $h_0$ is mainly set by SN+Cepheids and BAO+CMB, with the former strongly contributing to reduce the overall uncertainty. It must be stressed that in the $\eta$CDM model, at variance with standard $\Lambda$CDM, no $h_0$ tension exists, in that the determination from SN+Cepheids and BAO+CMB are consistent within $2\sigma$, with the BAO+CMB preferring slightly larger values. There is instead a tendency for CC+BAO$_{\perp}$ to give slightly smaller value of $h_0$ (though still marginally consistent within $3\sigma$), but the larger uncertainties on such dataset do not impact much the joint analysis. As to $\Omega_{m,0}$, the joint analysis is dominated by the constraint from BAO+CMB that tend to require an Universe with a small average curvature: on the one hand this does not permit too low values of $\Omega_{m,0}$, which is indeed larger than $\approx 0.3$ at $3\sigma$; on the other hand, the noise term acting in Eq.~(\ref{eq|kappa}) allows for values of $\Omega_{m,0}\approx 0.4$ appreciably smaller than $1$, at the price of enhancing the noise strength $\zeta$ or setting an appropriate negative value of the parameter $\alpha$ regulating the time dependence of the noise. In particular, the latter is mainly determined by BAO+CMB data that can actually probe, though in an integrated way, the conditions in the early Universe; the noise strength $\zeta$ is also appreciably constrained by SN cosmography as shown by Eq.~(\ref{eq|q}).

Recall from Sect. \ref{sec|basics} that the relative importance of noise and dilution terms in the mass-energy evolution equation (second of Eqs.~\ref{eq|friedrand}) is quantified by the ratio $t_\eta/t_{\rm exp}\sim 3/(\zeta^2\,h^{2\alpha-1})$; using the bestfit values of the noise and cosmological parameters, we can now evaluate that it is very large (noise is negligible) at high redshift, it crosses unity (i.e., noise starts to dominate and cosmic acceleration kicks in) at $z\approx 0.6$, and it amounts to $\sim 1/3$ at the present time.

\begin{figure}[!t]
\centering
\includegraphics[width=0.55\textwidth]{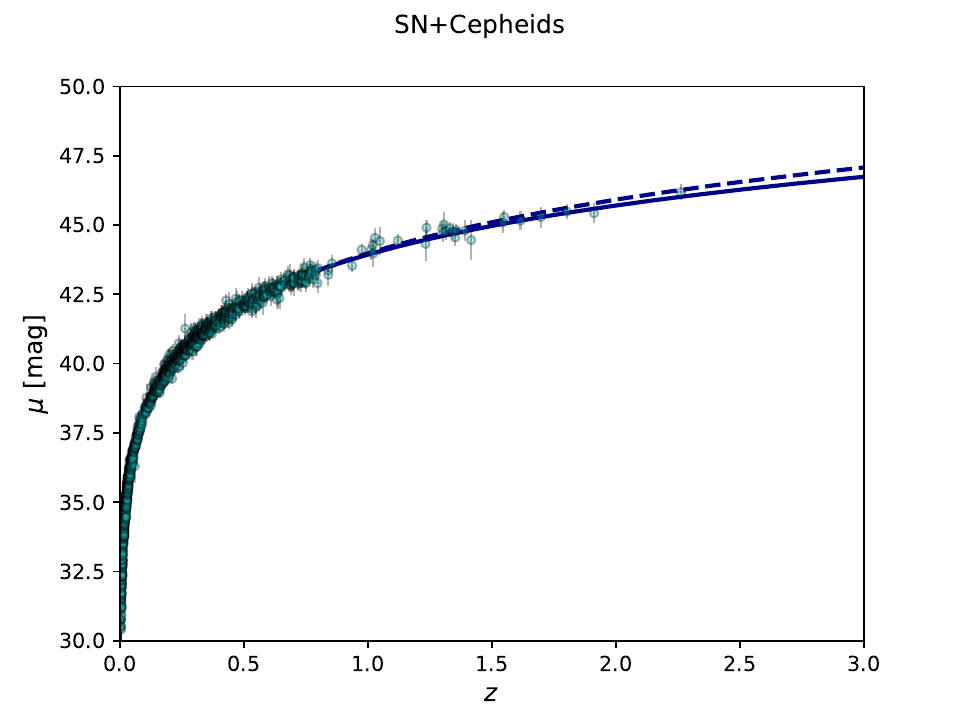}\\
\includegraphics[width=0.55\textwidth]{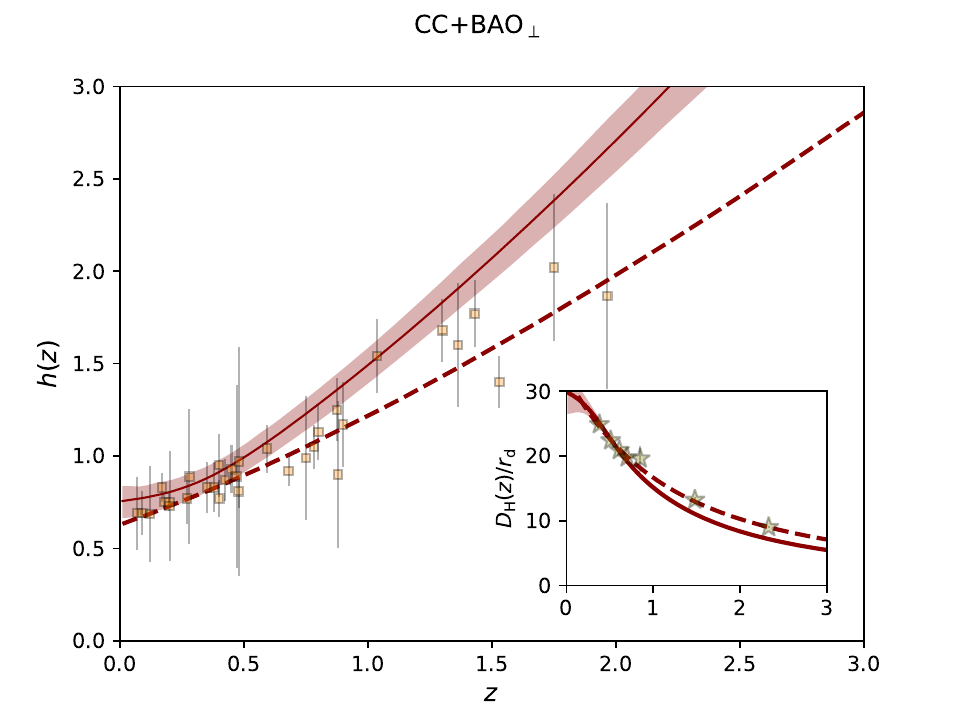}\\
\includegraphics[width=0.55\textwidth]{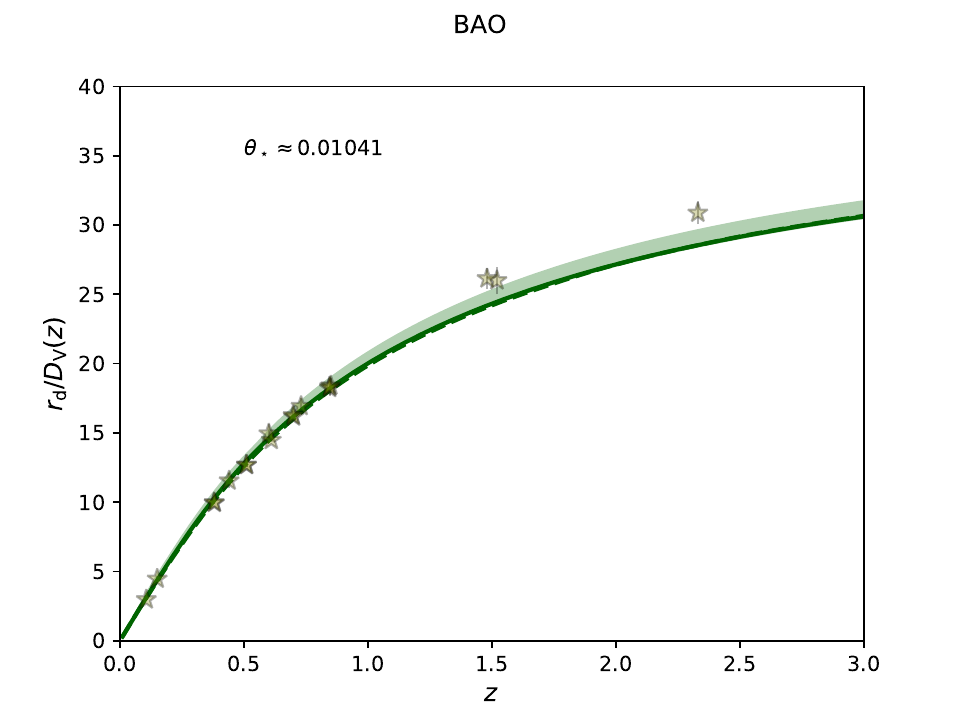}\\
\caption{Fits to SN+Cepheid (top panel), CC+BAO$_\perp$ (middle panel) and
BAO+CMB $\theta_\star$ (bottom panel) in the $\eta$CDM model. In each panel solid lines and shaded areas illustrate the median
and $2\sigma$ credible interval from sampling the posterior distribution
of the joint analysis, while the dashed line is the median from the posterior distribution of the
analysis to the individual dataset. Data and references for each observable are described in Sect. \ref{sec|cosmofit}.}\label{fig|obs}
\end{figure}

In Fig. \ref{fig|obs} we illustrate how the $\eta$CDM model performs on the fitted observables. In each panel the median (solid lines) and $2\sigma$ credible intervals (shaded areas) from sampling the posterior distribution of the joint analysis are shown; for reference, we also report (dashed lines) the median for the analysis of the individual observables. The individual fits are very good in all cases. The joint analysis fit performs decently on all the observables, being consistent with all the datapoints within $2\sigma$; the most evident discrepancy is with the determinations of the Hubble constant from CC+BAO$_{\perp}$, though the large errorbars of these datapoints impact marginally on the overall goodness of the fit. However, it should be stressed that the majority of the CC data reported in the Figure and exploited for the fit have been derived basing on the Bruzual \& Charlot (2003) Stellar Population synthesis libraries; it is known (see Moresco 2022) that using instead the Maraston \& Stromback (2011) models yields values of $H(z)$ systematically higher, especially toward higher $z$.

\begin{figure}[!t]
\centering
\includegraphics[width=0.525\textwidth]{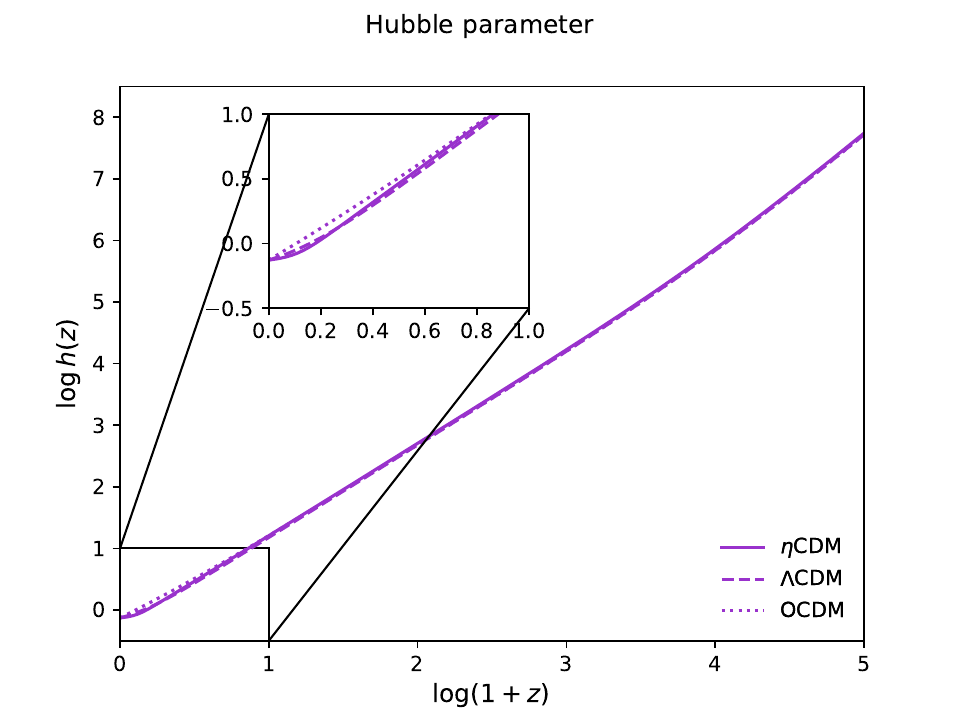}\\
\includegraphics[width=0.525\textwidth]{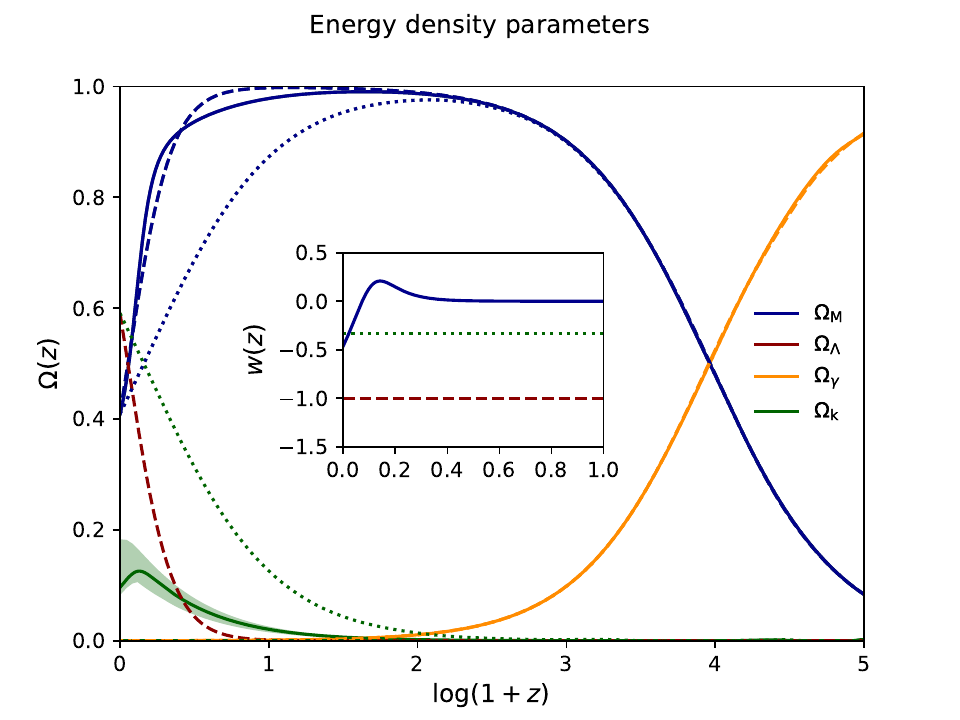}\\
\includegraphics[width=0.525\textwidth]{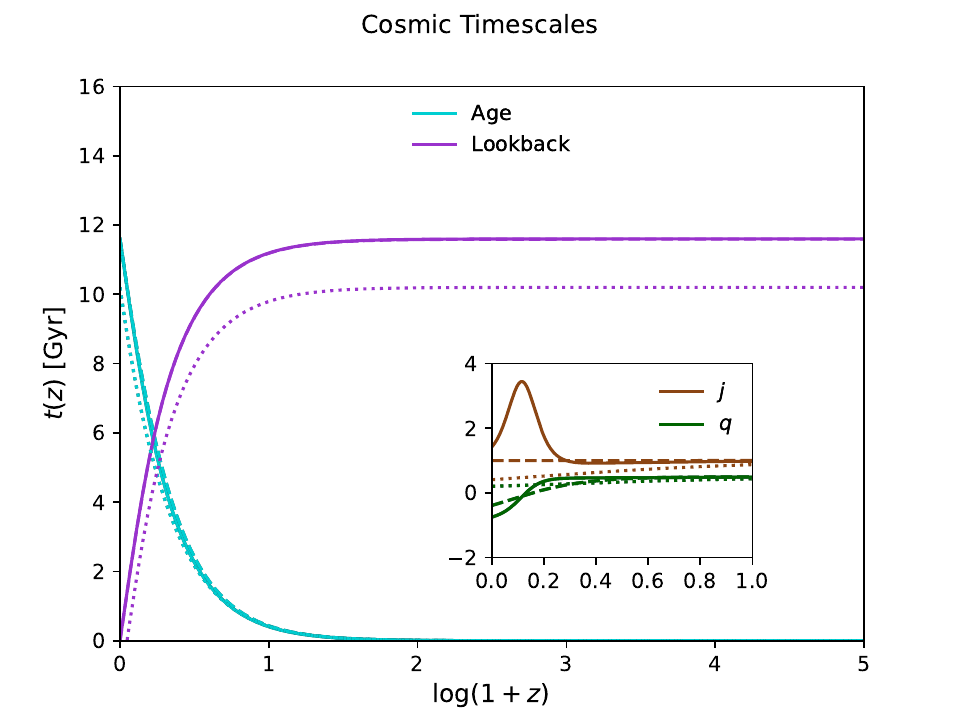}\\
\caption{Average evolution in the $\eta$CDM model. Top panel: Hubble parameter as a function of redshift; the inset zooms on the late-time evolution. Middle panel: energy density parameters of matter (blue), dark energy (red), radiation (orange), and curvature (green) as a function of redshift. The inset refers to the equation of state for the dominant energy density component at late times. Bottom panel: age of the Universe (cyan) and look-back time (magenta) as a function of redshift. The inset shows the late-time evolution of the deceleration parameter $q$ (green) and of the jerk parameter $j$ (brown).
In all the panels, solid lines are from $\eta$CDM with bestfit parameters $(h, \Omega_{m,0}, \zeta, \alpha)$ from the joint analysis of Sect. \ref{sec|cosmofit}, dashed lines for a flat $\Lambda$CDM with same parameters $(h, \Omega_{m,0})$ and dotted an open CDM model with same parameters $(h, \Omega_{m,0})$. In the middle panel, the green shaded area on the $\eta$CDM curvature parameter illustrates the typical $2\sigma$ uncertainty on the plotted quantities.}\label{fig|evo}
\end{figure}

In Fig. \ref{fig|evo} we illustrate the evolution of the bestfit $\eta$CDM Universe from the joint analysis; for reference, a flat $\Lambda$CDM and an open CDM model with the same values of $h_0\approx 0.75$ and $\Omega_{m,0}\approx 0.4$ are also reported for comparison. The top panel shows the evolution of the Hubble parameter, with the inset zooming in the late time Universe. The behavior of our $\eta$CDM model closely mirrors standard $\Lambda$CDM while progressively departing from an open model toward the present; this is because cosmic acceleration sets in with the right timing and the average curvature stays small despite the rather low matter density. The middle panel displays the evolution of the energy density parameters for the various components. At high redshift all models are indistinguishable since matter (blue) and radiation (orange) dominate the early-time evolution and the noise is negligible. Moving toward lower redshifts the open model departs from the other two since curvature (green) begins to grow early on, while $\eta$CDM and $\Lambda$CDM models are similar. Spatial flatness in $\Lambda$CDM is ensured by the presence of the additional dark energy component (red), while in $\eta$CDM the curvature is kept to low values (though not null) by the noise-induced drift term in Eq.~(\ref{eq|kappa}). In the middle panel, for reference the green shaded area on the $\eta$CDM curvature parameter illustrates the typical $2\sigma$ uncertainty on the energy densities. The inset focuses on the evolution of the equation of state parameter $\bar{w}$ for the dominant component at late times: an open universe is dominated by curvature with constant $\bar{w}=-1/3$; the $\Lambda$CDM model is dominated by dark energy with $\bar{w}=-1$ (generalization are obviously possible); the $\eta$CDM model is dominated by matter with an evolving $\bar{w}$ given by Eq.~(\ref{eq|wm}), which is zero at early times, grows slightly positive, attains a maximum and then decrease to negative values toward the present.

The bottom panel illustrate relevant cosmological timescales, namely the age of the Universe (cyan) and the lookback time (magenta) as a function of redshift; the $\eta$CDM and $\Lambda$CDM are almost indistinguishable, with an age approximately given by $1/H_0$, while the open model is clearly younger since it lacks the cosmic acceleration phase at late times. The inset displays the deceleration parameter $q$ and the jerk $j$. The former is positive for the open model which is decelerating, while assumes similar negative values around $q\approx -0.6$ for the accelerating $\eta$CDM and $\Lambda$CDM models. As for the jerk, in $\Lambda$CDM it is strictly unity, in open models is less than unity due to the presence of curvature, while for the $\eta$CDM model it has a non-trivial evolution at late-times due to the competition between curvature and noise-induced acceleration: first it grows above one, then attains a maximum value of a few and then decreases again toward current values slightly larger than unity. This non trivial behavior of the jerk could be a crucial observable to test the $\eta$CDM model in the near future when data from the \textit{Euclid} satellite will become available.

Finally, a caveat is in order. We have found that the bestfit values of $\Omega_{m,0}$ in the $\eta$CDM model amounts to around $0.4$, a value appreciably larger than the usual $\approx 0.3$ applying for $\Lambda$CDM. This will imply that,
at fixed power spectrum normalization (quantified by $\sigma_8$, the mass variance filtered on a scale of $8\, h^{-1}$ Mpc),
the number density of collapsed halos will be somewhat larger especially toward high $z$, an occurrence which may be pleasing in view of recent claims based on early JWST data (e.g., Labb\'e et al. 2023; Harikane et al. 2023; Xiao et al. 2023). In turn, the larger matter content will produce an enhanced lensing probability, leading to higher CMB lensing, galaxy-CMB lensing cross-correlation and cosmic shear amplitude. Actually all these observables probe only a combination of $\sigma_8$ and $\Omega_{m,0}$ to some power (e.g., for cosmic shear the relevant parameter is
$S_8\equiv \sigma_8\,\sqrt{\Omega_{m,0}/0.3}$), so that lowering somewhat $\sigma_8$ could still
met the associated observational constraints. Relatedly, in the present work we focused on the determination of a restricted set of basic cosmological parameters (essentially $\Omega_{m,0}$ and $h_0$) mainly via type I$a$ SN, BAO, and the position of the first peak in the CMB power spectrum. In the future, it will be necessary to conduct a global analysis on an extended cosmological parameter set (e.g., including primordial spectral index $n_s$, power spectrum normalization $\sigma_8$, etc.) by exploiting the overall CMB power spectra in intensity and polarization, and also including the aforementioned lensing observables; this could possibly slightly change the bestfit values of $\Omega_{m,0}$ and $h_0$ derived here.

\section{A (mildly) stochastic universe}\label{sec|stochasticity}

We now turn to quantify the random component of the $\eta$CDM model. According to the procedure outlined in Appendix \ref{sec|app_bridge} (see Eqs.~\ref{eq|diffbridge}), we need to solve the system of stochastic differential equations (of the Ito type)
\begin{equation}\label{eq|stocha}
\left\{
\begin{aligned}
\dot{\tilde{h}} &= -\cfrac{\tilde{h}}{\tau_0-\tau}+\cfrac{\zeta}{2}\,(\Omega_m+\Omega_\gamma)\, h^{\alpha+1}\, \eta(\tau)\\
\\
\dot{\tilde{\Omega}}_m &=-\cfrac{\tilde{\Omega}_m}{\tau_0-\tau}+\zeta\,(1-\Omega_m-\Omega_\gamma)\, \Omega_m\, h^\alpha\, \eta(\tau)\\
\\
\dot{\tilde{\Omega}}_\gamma &=-\cfrac{\tilde{\Omega}_\gamma}{\tau_0-\tau}+\zeta\,(1-\Omega_m-\Omega_\gamma)\, \Omega_\gamma\, h^\alpha\, \eta(\tau)~,
\end{aligned}
\right.
\end{equation}
with initial values $\tilde{h}(\tau_{\rm in})=\tilde{\Omega}_m(\tau_{\rm in})=\tilde{\Omega}_\gamma(\tau_{\rm in})=0$; we choose $\tau_{\rm in}$ corresponding to redshift $z_{\rm in}\approx 100$ but the results are unaffected as far as the $z_{\rm in}\gtrsim 10$ (i.e., a redshift so large that the noise term of the original process is negligible). In the above equations
recall that by definition $h=\bar{h}+\tilde{h}$ and $\Omega_{m,\gamma}=\bar{\Omega}_{m,\gamma}+\tilde{\Omega}_{m,\gamma}$ hold, with the barred quantities constituting the solution for the average behavior found in the previous sections.

\begin{figure}[!t]
\centering
\includegraphics[width=0.45\textwidth]{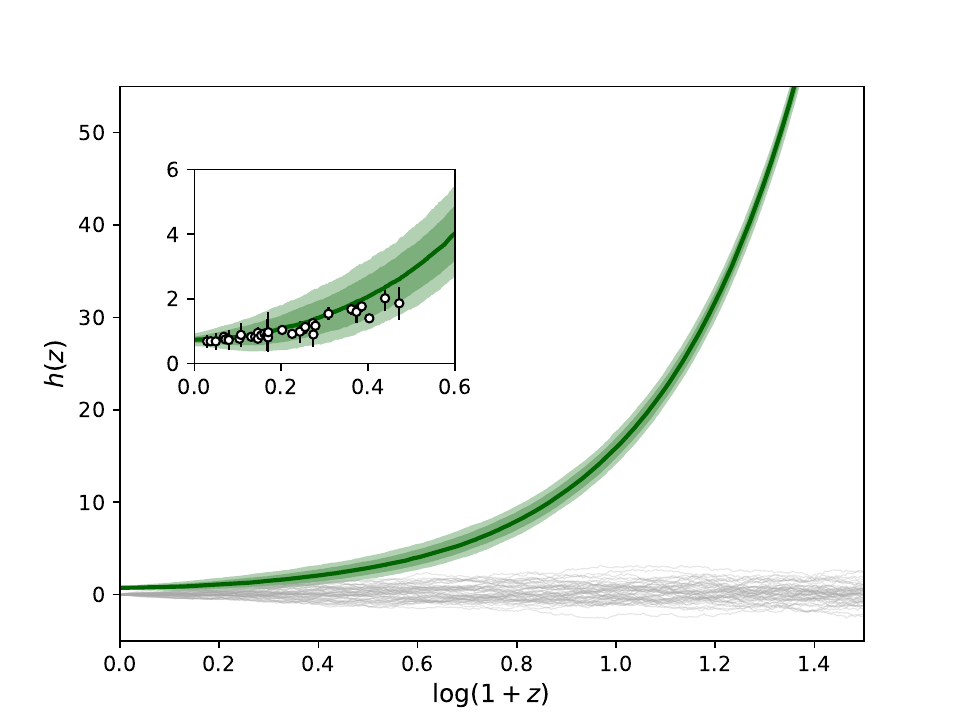}
\includegraphics[width=0.45\textwidth]{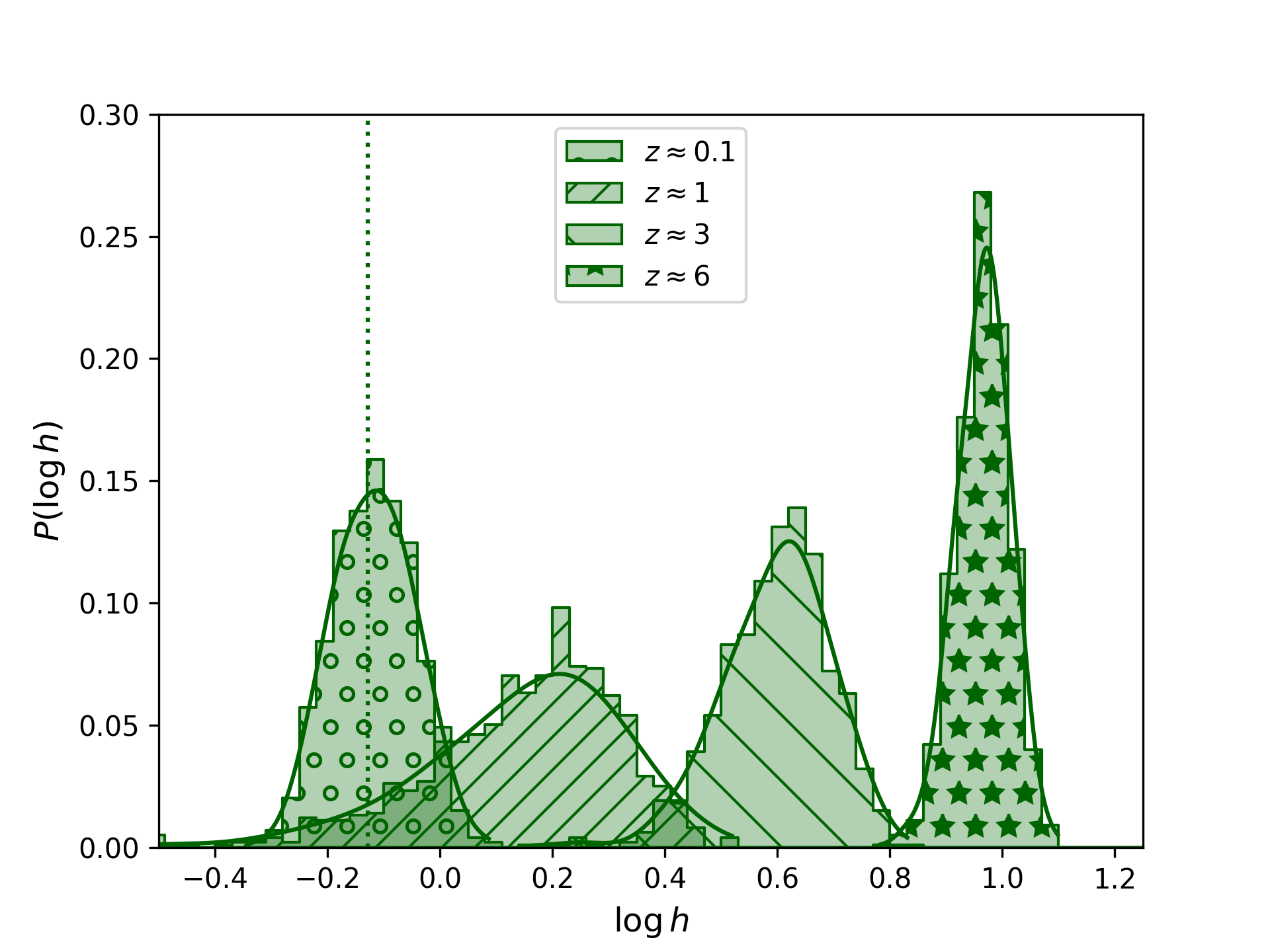}\\
\includegraphics[width=0.45\textwidth]{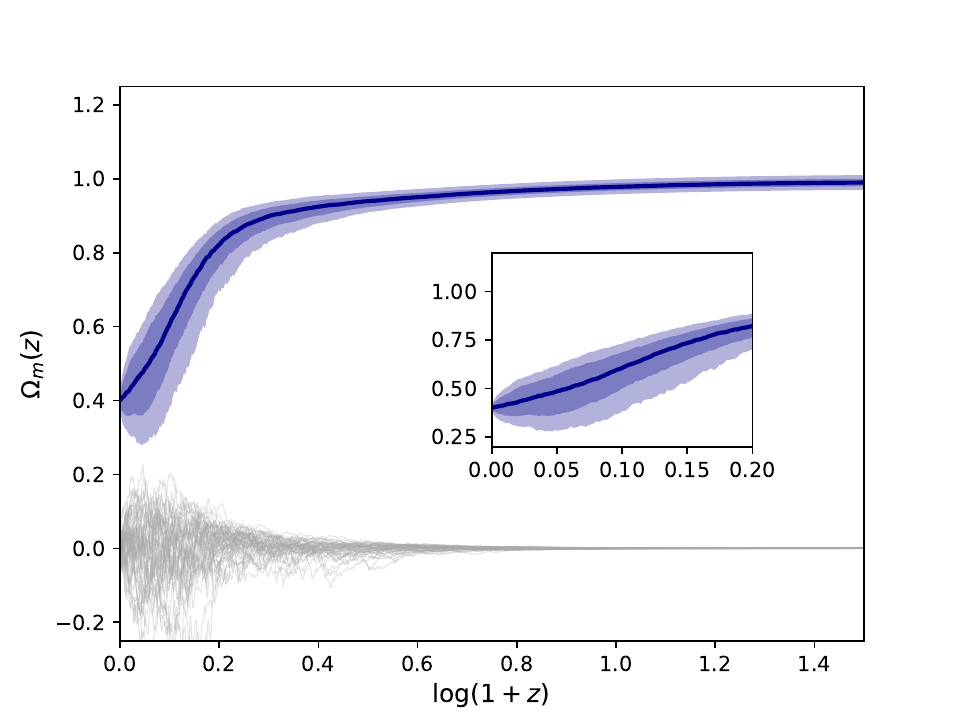}
\includegraphics[width=0.45\textwidth]{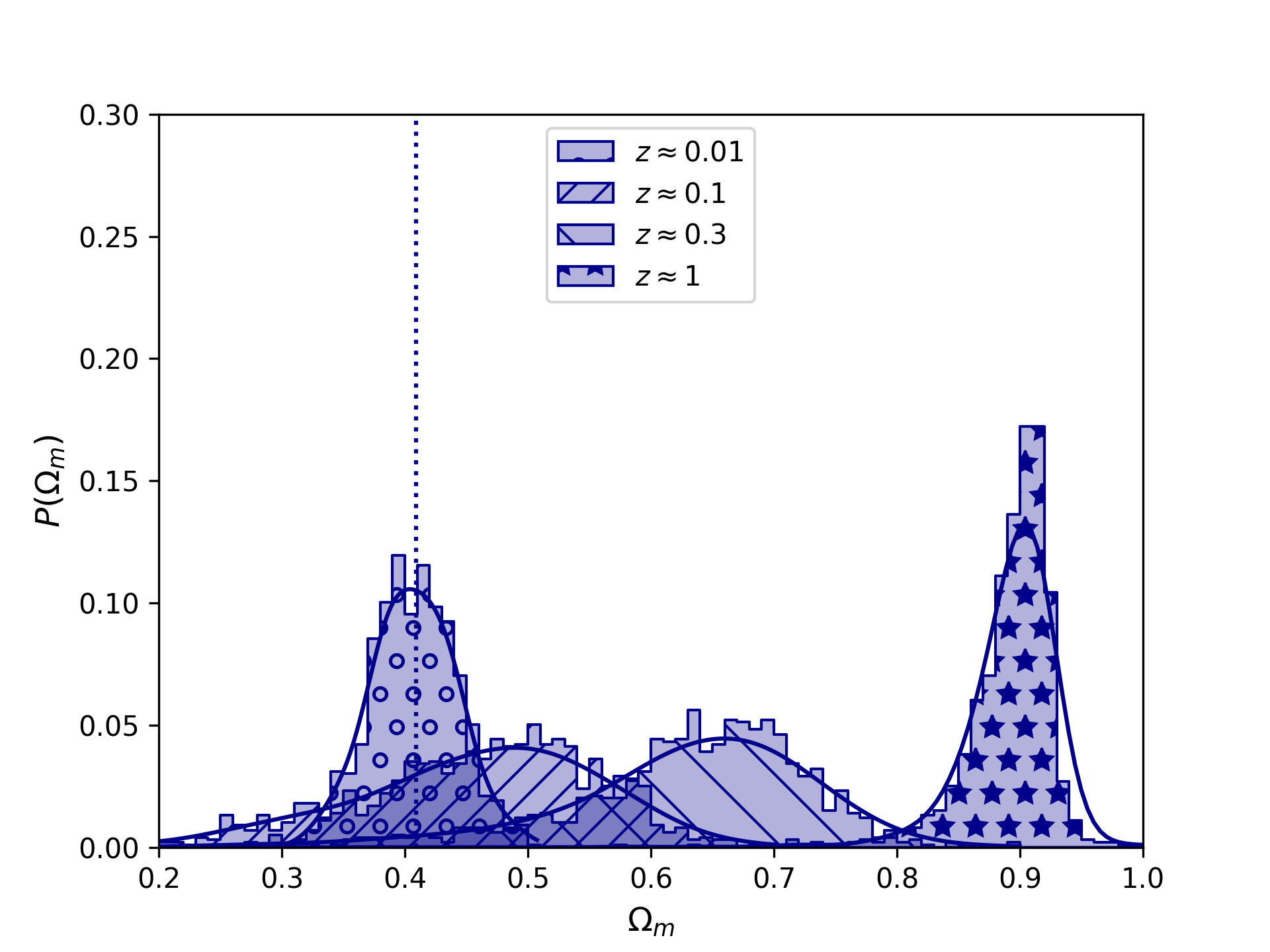}\\
\includegraphics[width=0.45\textwidth]{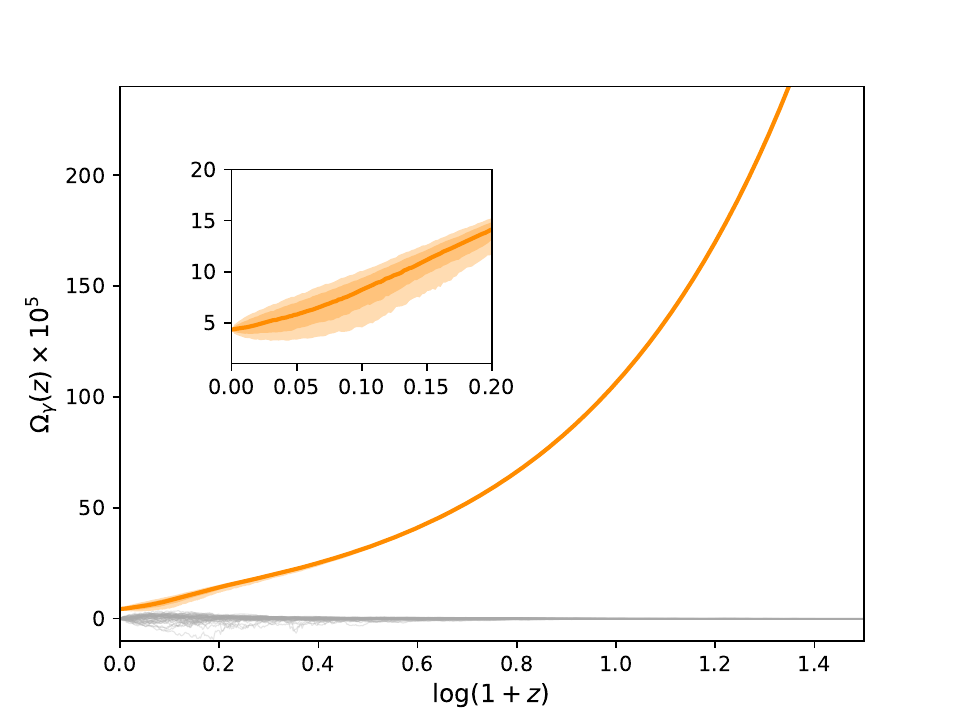}
\includegraphics[width=0.45\textwidth]{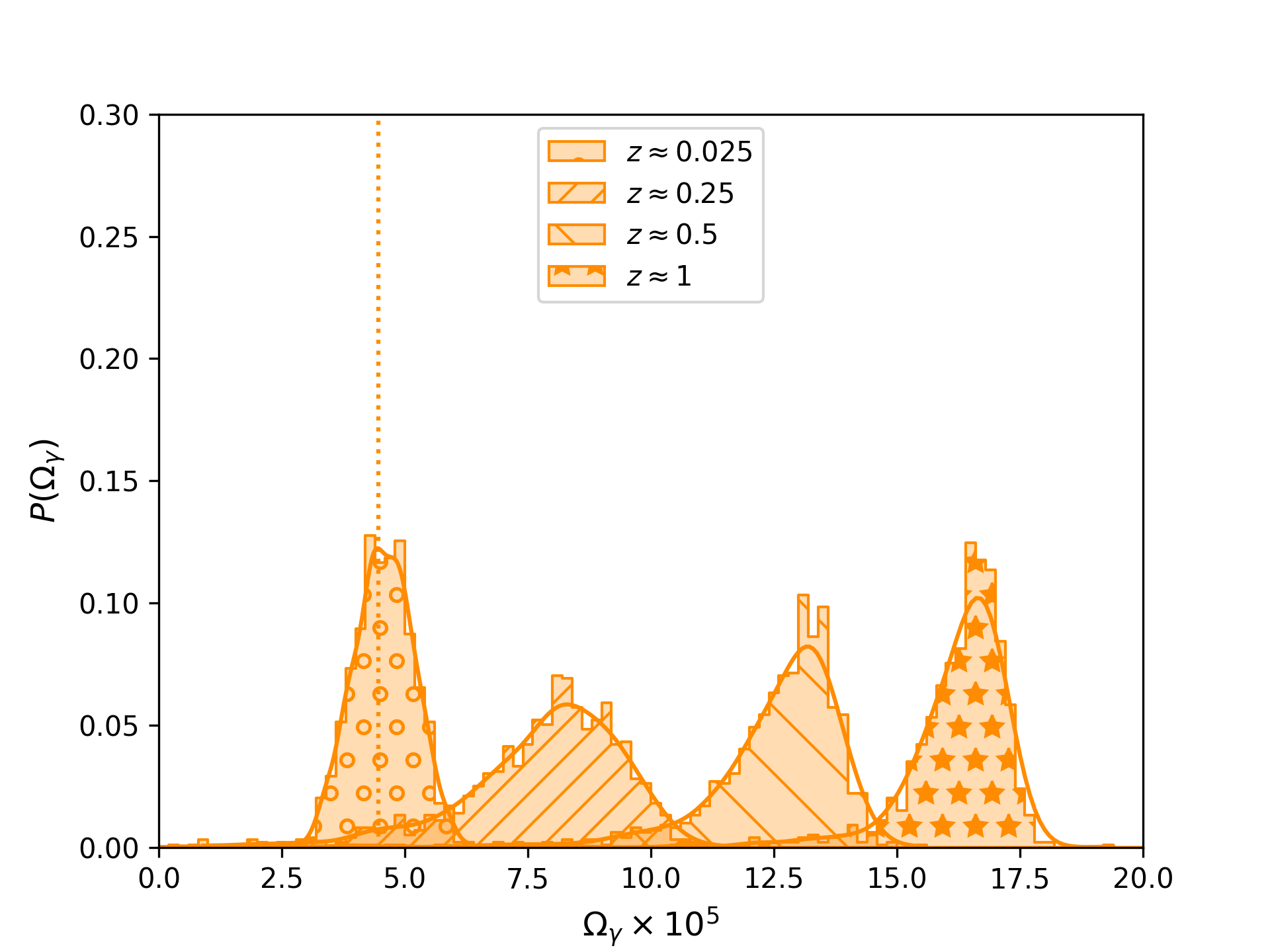}\\
\caption{Cosmic evolution and probability distribution of the Hubble parameter (top panels), matter density parameter (middle panel) and
radiation density parameter (bottom panel) in the $\eta$CDM model. Left panels illustrate the average evolution (colored solid line), the $1-2\sigma$ variance of the different quantities as originated by the noise (colored shaded areas), and some examples of the random component depurated by the average (grey lines); the insets zoom on the late-time evolution, with that for $h(z)$ also reporting data from cosmic chronometers. The right panels illustrate the corresponding probability distributions (histograms and kernel density estimators as solid lines),
at different representative redshifts in the late Universe (see legend); the dotted vertical line shows the present value (see Sect. \ref{sec|cosmofit}).}\label{fig|stocha}
\end{figure}

We solve the above system with a Euler-Maruyama method\footnote{The Euler-Maruyama method is a generalization of the Euler method for stochastic differential equations. It consists in discretizing a Ito-type equation $\dot x=f(x)+g(x)\,\eta(t)$ as $x(t_{j+1})=x(t_j)+f(t_j)\,(t_{j+1}-t_j)+g[x(t_j)]\,\sqrt{t_{j+1}-t_j}\,w_j$ in terms
of random weights $w_j\in \mathcal{N}(0,\sqrt{2})$ extracted from a normal distribution with zero mean and variance two (e.g., Kloeden \& Platen 1992; Risken 1996).}, and determine the stochastic evolution of $\tilde{h}$ and $\tilde{\Omega}_{m,\gamma}$ by running $10^4$ realizations of the stochastic process. Then we combine such random components with the average ones from Sect. \ref{sec|deterministic} to reconstruct the overall cosmic history of the Hubble $h$ and energy density parameters $\Omega_{m,\gamma}$ in each patch of the Universe. For an observer measuring the current values of these quantities as estimated in Sect. \ref{sec|cosmofit} by comparison with cosmological observables, the overall evolution is plotted in Fig. \ref{fig|stocha}  (in the overall variance we also include the uncertainties in the determination of $h_0$ and $\Omega_{m,0}$ from cosmological observables, but this is negligible with respect to that induced by the noise). Specifically, there we illustrate the mean value and the $1-2\sigma$ variance induced by the noise as a function of redshift. In the right panels we also display histograms showing the probability distribution of these quantities at some representative redshifts in the late Universe when noise is appreciably active; these represents the spatial distribution of the cosmological quantities among different patches of the Universe.

The Hubble parameter $h$ fluctuates such that its $1\sigma$ variance is about $0.07$ dex at $z\approx 0.1$, increases up to $0.13$ dex around $z\approx 1$ and then starts to decrease amounting to $0.08$ dex at $z\gtrsim 3$ and becoming progressively negligible at higher redshifts. For $\Omega_m$, the evolution of the variance is faster, being $0.03$ dex at $z\approx 0.01$, increases up to $0.08$ dex
at $z\approx 0.1-0.3$ and then quickly decreases to $0.03$ dex at $z\approx 1$ and to negligible values at higher $z$; fluctuations in $\Omega_\kappa$ are similar to these. For $\Omega_\gamma$
typical values of the variance are $0.05$ dex at $z\approx 0.025$, a maximal $0.1$ dex at $z\approx 0.25$, and $0.03$ at $z\approx 1$.
As already explained, such fluctuations of the relevant cosmological quantities should be interpreted as residual deviations from isotropy/homogeneity on different patches of the Universe. It is a specific prediction of the $\eta$CDM model that one should observe the anisotropies/inhomogeneities estimated here via a tomographic analysis in redshift shells on different patches of the Universe. Given the magnitude of the fluctuations, such observations will be quite challenging though possibly within the reach of future wide galaxy surveys like \textit{Euclid} or \textit{LSST}.

As anticipated in Sect. \ref{sec|basics} one can now estimate the typical coarse-graining scale associated to the $\eta$CDM model. Given that the fluctuations of the overdensity field $\log \Omega\sim \log (1+\delta)$ induced by the noise amounts to $\lesssim 0.08$ dex, we can exploit the relation between the smoothing scale and variance in the overdensity field from Fig. \ref{fig|Nbody} (bottom right panel) and its accuracy of $\lesssim 10\%$ (see Repp \& Szapudi 2017, 2018) to estimate an $\eta$CDM coarse-graining scale of around $42\pm 5\, h^{-1}$ Mpc. On the one hand, such an estimate must be taken with care, since it has been derived basing on $N-$body simulations performed in $\Lambda$CDM; on the other hand, the relation between smoothing scale and variance in the overdensity field should not vary much given that $\Lambda$CDM and $\eta$CDM have very similar (spatially-averaged) evolution in cosmic time. All in all, we can confidently (at $2\sigma$) quote for $\eta$CDM a coarse-graining scale in the range $30-50\, h^{-1}$ Mpcs. Pleasingly, the coarse graining scale so obtained corresponds to the typical size of the structures present in the cosmic web, as envisaged in building our stochastic framework in Sect. \ref{sec|basics}. Furthermore, on patches of this size, the corresponding fluctuations in matter density $\log \rho\sim \log (\Omega\, h^2)$ are of order $\lesssim 0.1$ dex, implying in Eq.~(\ref{eq|friedrand}) minor deviations from local energy conservation induced by the noise term. Finally, the mean bulk motions on such scale is expected to be around $\sim 300$ km s$^{-1}$, which is substantially smaller than the Hubble flow $\sim 3500$ km $s^{-1}$, so ensuring that the Friedmann equation can still constitute a safe approximation.

\section{The fate of the Universe}\label{sec|destiny}

What will be the fate of the Universe in the $\eta$CDM model? To answer the question, we go back to Eqs.~(\ref{eq|friednew}); in particular, since radiation is negligible at late times (and even more
in the future), one can focus on the first two equations for the evolution of $\bar h$ and $\bar{\Omega}_m$. It is quite easy to realize,
and straightforward to confirm numerically, that the system of equations admits an
attractor solution in the infinite future; this can be found by putting $\dot{\bar{h}}=\dot{\bar{\Omega}}_m=0$ and solving for the
asymptotic values $h_\infty$ and $\Omega_{\infty}$. A simple algebraic calculation yields:
\begin{equation}\label{eq|destiny}
\left\{
\begin{aligned}
& 3\, \Omega_{\infty}^2+2\,(4-\alpha)\,\Omega_{\infty}-4 = 0~,\\
\\
& h_\infty = \left[\zeta^2\,\left(1-\cfrac{4-\alpha}{2}\, \Omega_\infty\right)\right]^{1/(1-2\alpha)}~.
\end{aligned}
\right.
\end{equation}
Therefore $\Omega_m$ tends to a constant, nonzero value that depends solely on the parameter $\alpha$ describing the time dependence of the noise;
for the value $\alpha\approx -3/2$ as estimated from the comparison with cosmological datasets in Sect. \ref{sec|cosmofit}, one finds $\Omega_\infty\approx 1/3$. Moreover, from Eq.~(\ref{eq|wm}) it follows that the effective equation of state for matter asymptotes to $w_\infty\approx -1$; as a matter of fact, in the infinite future matter will behave like a cosmological constant!
The limiting value of $h$ depends also on the noise strength $\zeta$, and for the bestfit parameters estimated in Sect. \ref{sec|cosmofit} one finds $h_\infty\approx 0.7$; clearly from Eq.~(\ref{eq|q}) it follows that $q_\infty\approx -1$. This implies that the Universe will continue to expand forever in an exponential way so that the asymptotic behavior of the scale factor $a(t)\propto e^{h_\infty\, t}$ will apply. Furthermore Eq.~(\ref{eq|kappa}) directly implies that $\bar{\Omega}_\kappa$ will tend to zero as $\bar{\Omega}_\kappa\propto a^{-3\zeta^2\,h_\infty^{2\alpha-1}\,\Omega_\infty^2/2}\rightarrow 0$ in the infinite future.

A crucial remark is in order here. The current estimates of the parameters $h_0\approx 0.75$ and $\Omega_{m,0}\approx 0.4$ inferred from comparison with cosmological datasets in Sect. \ref{sec|cosmofit} are very close to the asymptotic values $h_\infty$ and $\Omega_\infty$ that will be attained in the infinite future. This occurrence relieves coincidence problems even in a wide sense; besides waiting for conditions in the Universe that are suitable for the development of life, there is no coincidence at all in the fact that we currently measure certain values of the cosmological parameters $h_0$ and $\Omega_{m,0}$. This is because the Universe will have values very close to those for essentially an infinite amount of time. In other words, as soon as the noise starts affecting the cosmic evolution and brings cosmic acceleration into the game, such parameters will stay put or change very little.

This is at variance with the situation in $\Lambda$CDM, where in the past $\Omega_m$ was close to $1$, $\Omega_\Lambda$ was close to zero and $h$ very large, while in the infinite future $\Omega_m$ will tend to zero, $\Omega_\Lambda$ to $1$, and $h$ to $\sqrt{\Omega_{\Lambda,0}}\, h_0$. Given that the currently measured values are in between these extremes (even more, $\Omega_m$ and $\Omega_\Lambda$ are of the same order), the coincidence problem is very pressing for $\Lambda$CDM: why do we live in this very precise moment of the cosmic history?
As shown above, the issue is cleared in $\eta$CDM.

\section{Frequently asked questions}\label{sec|discussion}

Below we try to answer some questions that may arise in connection with our non-standard $\eta$CDM cosmological model.

\begin{itemize}

\item \emph{What about the meaning of the noise term?}

The meaning of the noise term is to provide a mean-field description of the slightly different evolution for patches of the Universe that are tens Mpcs in size, which are enforced by local inhomogeneities, matter flows due to anisotropic stresses, tidal forces, gravitational torques, and other complex gravitational
processes that for all practical purposes are extremely difficult to model ab-initio or to handle (semi-)analytically.

The situation is somewhat analogous to the classic description
of Brownian motion: a microscopic particle immersed in a
fluid continuously undergoes collisions with the fluid molecules;
the resulting motion, despite being deterministic, appears
to be random at the macroscopic level, especially to an external
observer who has no access to the exact positions and velocities
of the innumerable fluid molecules and to the initial conditions
of the particle. In the way of a statistical macroscopic description, the
problem is effectively treated via a stochastic differential
equation driven by a fluctuating white noise, which allows to
implicitly account for the complex microscopic dynamics of
the system.

Note that the system's state often influences the
intensity of the driving noise, like when the Brownian
fluctuations of a microscopic particle near a wall are reduced by
hydrodynamic interactions, so that the noise becomes multiplicative
in terms of a nonuniform diffusion coefficient. This adds complexity to the
dynamics because the multiplicative nature of the noise
brings a noise-induced drift into the game, which can appreciably affect
the overall evolution of the system (even its average component).
Similar stochastic models with multiplicative noise have been
employed to describe a wide range of physical phenomena,
from Brownian motion in inhomogeneous media or in close
approach to physical barriers, to thermal fluctuations in
electronic circuits, to the evolution of stock prices, to computer
science, to the heterogeneous response of biological systems
and randomness in gene expression (e.g., Risken 1996;
Mitzenmacher 2004; Reed \& Jorgensen 2004; Paul \& Baschnagel 2013). In cosmology,
a similar formalism, though with different premises and aims,
has been also exploited in models of stochastic inflation
(see Vilenkin 1983; Starobinski 1986; Nakao et al. 1988; Salopek \& Bond 1990; see recent review
by Cruces 2022 and references therein) and in the field of structure formation to
predict the halo mass function and related statistics (e.g., Bond et al. 1991; Mo \&  White 1996;
Lapi \& Danese 2020; Lapi et al. 2022).

In the $\eta$CDM model, the noise-induced drift
is associated to the multiplicative nature of the stochastic term in the
mass-energy evolution equation. As we have shown, it
can substantially affect the cosmic dynamics at late times,
driving an accelerated expansion, forcing matter to behave as a negative
pressure component, and keeping the curvature to small values even in a low
density, matter-dominated Universe.

\item \emph{What about the modeling of the noise term?}

In the modeling of complex systems (see references above), the stochastic equations and the noise terms are
designed on purpose to effectively describe the macroscopic dynamics, and then
checked by comparison with observations and/or numerical simulations.

In the context of the $\eta$CDM model, the noise term can be naively justified by the fact that at given cosmic time the overdensity $\log (1+\delta)\propto \log \rho$ smoothed on a scale of tens Mpcs in size is expected to follow a lognormal distribution; as shown in Fig. \ref{fig|Nbody}, this is proved by simulations to be a good approximation in $\Lambda$CDM (see Reep \& Szapudi 2018; also Kayo et al. 2001) and there are theoretical arguments supporting this assumption in general (see discussion by Coles \& Jones 1991; also Neyrinck et al. 2009 and Repp \& Szapudi 2018).
In terms of basic stochastic processes, an ensemble of regions whose density evolves stochastically in time under a Gaussian white noise $\eta(t)$, or in other words for which ${\rm d}_t\log \rho\sim \dot \rho/\rho\propto \eta(t)$ features as solution a
lognormal distribution with time-dependent variance, as recalled in Appendix \ref{sec|app_stocha} (see also Risken 1996; Paul \& Baschnagel 2013).
Although the equations ruling the $\eta$CDM model (cf. Eqs.~\ref{eq|friedrand}) describe a more complex stochastic system, this analogy has inspired us to model the noise term as $\dot \rho\sim \zeta\,\,H^\alpha\, \rho$, with the parameters $\zeta$ and $\alpha$ describing our ignorance on the present value and on the redshift evolution of the variance in the density distribution for a generic cosmology that can be in principle different from $\Lambda$CDM.

The adopted modeling for the noise also avoids adding too much complexity (or parameters)
and yet satisfying a few physical requirements.
First, the stochasticity is driven by a Gaussian white noise. In the absence of a detailed control
on the gravitational dynamics this is the natural choice in the modeling of stochastic systems;
in future developments such an assumption can be relaxed by allowing for more complex frameworks with correlated or fractional noise, etc.
Second, the linear dependence of the noise on the energy density of cosmic components pleasingly does not break
the linear nature of the mass-energy evolution equation.
Third, the inverse dependence of the noise term on the Hubble parameter guarantees that
the noise term vanishes in the early Universe,
where (statistical) isotropy/homogeneity is robustly verified via CMB observations.
The choice of a power-law dependence is somewhat arbitrary and
has been dictated by our intention of keeping the treatment as simple as possible and
limiting the number of noise-related parameters. However, we have checked that the implications on the cosmic dynamics are quite robust against different parameterization of this dependence (we tested, e.g., an exponential function of $h$), though making the equations less transparent. Thus in this first investigation we prefer to avoid such complications.

All in all, to describe the phenomenon of the cosmic acceleration the $\eta$CDM model features
the same number of parameters of the standard $\Lambda$CDM cosmology. In $\Lambda$CDM, an additional component is added
with abundance $\Omega_{\Lambda,0}$ and equation of state $w_\Lambda$ (whose evolution
in turn can be characterized by one or more parameters).
In $\eta$CDM no additional component is added and
two parameters are needed to describe the strength and redshift dependence of the noise; the ensuing evolution of the cosmic dynamics at late times is completely specified (also including an effective, time-dependent equation of state
for the matter component, cf. Eq.~\ref{eq|wm}).

\item \emph{What about alternative physical interpretations of stochasticity on large scales?}

In the past literature other origins for stochasticity on cosmological scales, though rather different from our viewpoint, have been envisaged; for the sake of completeness we briefly mention these below.

Stochasticity can be plausibly originated by baryonic physics ongoing in collapsed objects (e.g., galaxies or clusters), such as turbulence, feedback processes from supernova explosions or AGN outburts/jets, etc. The problem with this scenario is that such effects are rather contained in space, influencing patches from kpc to at most Mpc scales. In fact, it has been argued that the effective causal limit of a galaxy is set by scales attained by matter flows around it over the age of the Universe, rather than by its usual light cone (see Ellis \& Stoeger 2009). However, there are some proposed mechanisms to transfer such randomness on much larger cosmological scales via chaotic dynamics or spontaneous stochasticity (a kind of `butterfly effect'), but definite conclusions on their effectiveness are still far from being drawn (see Neyrinck et al. 2022 and references therein).

Stochasticity is also naturally originated at the quantum level. In this vein, a spatially fluctuating field that describes random matter/radiation creation or disappearance could permeate the Universe (e.g., Sivakumar et al. 2001; Lima et al. 2008; Amin \& Baumann 2016; Mantinan et al. 2023). Albeit being highly exotic, the problem with such a scenario is that it requires some unspecified mechanism (a sort of quantum spontaneous stochasticity, see Eyink \& Drivas 2015) to allow such tiny fluctuations to expand, reinforce and become large enough to ultimately influence astrophysical and cosmological scales; moreover, a specific coupling of such a random field with the gravitational metric is possibly required to affect the cosmic dynamics at late times.

\item \emph{Is the $\eta$CDM model violating the Cosmological or the Copernican principles?}

The Cosmological principle \emph{is} violated, but \emph{with a grain of salt}. The $\eta$CDM model does not suggest strong violation of isotropy or homogeneity on horizon scales, that would require to change completely our view of the Universe and the description of the gravitational metric (e.g., reverting to a Lemaître–Tolman–Bondi Universe or a Swiss-cheese Universe; e.g., see Marra et al. 2007). It instead just advocates, as supported by numerical simulations (see Sect. \ref{sec|basics}), that small deviations of isotropy/homogeneity are present on scales of tens Mpc associated to the quasi-linear structures of the cosmic web.

The Copernican principle \emph{is not} violated, nor in the strict (humans on Earth are not privileged observers) nor in the enlarged
(no one in the Universe is a privileged observer) sense. In the $\eta$CDM violation of isotropy and homogeneity are small and imply minor fluctuations of the cosmological quantities on top of a still dominant, yet noise-informed average evolution. Every observer in the Universe will measure similar values of the cosmological parameters and should be able to statistically verify the same small deviations from isotropy/inhomogeneities on large scales.

\item \emph{What is the origin of the cosmic acceleration in $\eta$CDM?}

In $\eta$CDM the cosmic acceleration is caused by the noise-induced drift appearing in Eqs.~(\ref{eq|friednew}), which represents the ensemble-averaged evolution of the stochastic system described by Eq.~(\ref{eq|friednormrand}). It is instructive to mathematically and physically understand how the cosmic acceleration comes about from a heuristic analysis of the original Friedmann, mass-energy evolution and acceleration equations (see Eqs.~\ref{eq|friedrand} and \ref{eq|friedacc}). This can be done on considering the following argument based on simple scaling laws: neglecting curvature from the Friedmann equation $H\sim \rho^{1/2}$ applies, hence the random term in the mass-energy evolution equation scales as $\dot\rho\sim \zeta\,\rho\, H^\alpha\, \eta\sim \zeta\, \rho^{1+\alpha/2}\,\eta$, while that in the acceleration equation goes like $\ddot a\sim \zeta\,\rho\,H^{\alpha-1}\, \eta \sim \zeta\, \rho^{(1+\alpha)/2}\,\eta$; given that $\alpha\sim -1.5$ the scaling exponent $1+\alpha/2>0$ in the equation for $\dot\rho$ is positive, while the one $(1+\alpha)/2<0$ in the equation for $\ddot a$ is negative.

This implies that a fluctuation induced by the noise $\eta$ will be reinforced in high-density regions (e.g., filaments, knots) as for the variation of $\rho$, but at the same time it will be softened in terms of contribution to the acceleration; contrariwise, in low-density regions (voids) the fluctuation will be damped in terms of $\dot\rho$ while it will be amplified in terms of $\ddot a$. All in all, this mechanism would cause statistically the overall ensemble of patches to drift toward an evolution dominated by low-density regions, and characterized by an enhanced expansion rate.

\item \emph{Is $\eta$CDM a backreaction model?}

Our stochastic cosmology is inspired and has some features in common with Newtonian backreaction models, so it is worth pointing out the similarities and crucial differences between these frameworks.

Newtonian (also called kinematical) backreaction is a class of models originated in a seminal paper by Buchert \& Ehlers (1997), who explored the effects of matter anisotropy/inhomogeneities on the expansion rate for a given patch of the Universe. Specifically, the latter authors perform spatial averaging on an arbitrary domain of volume $V$ to derive an equation for the evolution of the local expansion factor $a\equiv V^{1/3}$; working in a Newtonian framework and assuming mass conservation and a pressureless matter component, they get the modified acceleration equation
\begin{equation}\label{eq|backreact}
\frac{\ddot{a}}{a} = -\frac{4\pi\, G}{3} \langle\rho_m\rangle +\frac{2}{9}\, (\langle \dot a^2\rangle-\langle \dot a \rangle^2)+\frac{2}{3}\langle \omega^2-\sigma^2\rangle~,
\end{equation}
where $\langle\rho_m\rangle$ is the average density in the volume, while $\omega$ and $\sigma$ are the magnitude of the rotation and shear tensors. The terms on the right hand side account for the fact that anisotropy/inhomogeneities in the matter distribution can modify the average expansion rate of the volume, which for particular conditions could also be driven to accelerate (e.g., Kolb 2011; Buchert \& Rasanen 2012). However, backreaction effects depends on the size of the volume under consideration, and it is generally assumed that these rapidly decrease for scales larger than the largest inhomogeneity (see Kaiser 2017); on the other hand, this treatment neglects the possibly different evolutions of the various patches because of the local inhomogeneities/anisotropies, matter flows, and the overall consequence of sampling effects on the cosmic dynamics.

The $\eta$CDM framework shares with the above backreaction models the general idea that structure formation could possibly modify the expansion rate on larger scales. However, it attacks the problem with a different statistical approach, that envisages the Universe tessellated with patches of tens Mpc size where residual anisotropy/inhomogeneity, as indicated by numerical simulations, is still present. Such regions will undergo slightly different evolutions due to local inhomogeneities, matter flows, and to many complex gravitational processes. The detailed dynamics is extremely difficult to follow (semi-)analytically, hence a statistical description for the evolution of the different patches is adopted in terms of a stochastic noise term in Eq.~(\ref{eq|friedacc}), whereas the global evolution of the Universe is then derived by averaging the behavior over the patch ensemble.

In this respect our approach reminds of the AvERA algorithm (Average Expansion Rate Approximation; see Racz et al. 2017); this is a procedure to extract the cosmological expansion rate from a $N-$body simulations via a volume averaging technique. The basic idea of AvERA is to account for local inhomogeneities by inverting the usual order of volume-averaging and expansion rate computation in the simulation: first small patches of the Universe with a certain coarse graining scale are evolved via the standard Friedmann equation (with no dark energy) according to the local density, and then the global expansion rate is obtained by volume-averaging over the local scale-factor increments; remarkably, the net outcome is a global expansion history mirroring that of the standard $\Lambda$CDM model.

In AvERA and in the $\eta$CDM model the cosmic acceleration originates by the same basic underlying physics, i.e. that the nature of the large-scale structure formation places far more volume in underdense than in overdense regions, causing the average expansion to skew toward acceleration. Both in the AvERA and in the $\eta$CDM model the coarse-graining scale involved is essentially a free-parameter
(a spatial smoothing scale used in the simulation for AvERA, and a quantity fully specified by the noise parameters $\zeta$ and $\alpha$ for the $\eta$CDM model), that is set by comparison with data.

However, in AvERA the coarse-graining scale is found to be smaller than a few Mpc (which correspond to $\lesssim 10^{12}\, M_\odot$) while in the $\eta$CDM model it turns out to be of several tens Mpc (which are associated to the quasi-linear structures of the cosmic web). This difference in the effective coarse-graining scale can be traced back to the diverse assumptions of the two models. Specifically, the AvERA approach is rooted on the separate Universe conjecture, meaning that spherically-symmetric patches of the Universe are assumed to behave like isolated islands evolving with their own energy density $\Omega \sim 1+\delta$, while anisotropic stresses, tidal forces, external environment, cross-talks of different regions by flows of matter and radiation are neglected (see also Buchert 2018). On the other hand, the $\eta$CDM model allows for such processes to occur, admittedly at the price of reverting to a mean-field statistical description in terms of a phenomenological yet physically reasonable noise term.

To sum up, the AvERA and the $\eta$CDM constitute somewhat complementary approaches, concurring to suggest that structure formation on non-linear and quasi-linear spatial scales can have relevant effects on the overall cosmic expansion.

\end{itemize}

\section{Summary}\label{sec|summary}

We have proposed a new model of the Universe called $\eta$CDM. Its marking feature is a controlled stochastic evolution of the cosmological quantities, that is meant to render the effects of small deviations from homogeneity/isotropy on large scales of tens Mpc size at late cosmic times, associated to the emergence of the cosmic web. Specifically, we prescribe that, still in the context of standard general relativity, the evolution of the matter/radiation energy densities in different patches of the Universe can be effectively described by a stochastic version of the mass-energy evolution equation.
The latter includes, besides the usual dilution due to cosmic expansion, an appropriate multiplicative and time-dependent noise term that statistically accounts for local fluctuations due to inhomogeneities and matter flows induced by anisotropic stresses and many complex gravitational processes. The different evolution of the patches as a function of cosmic time is rendered via the diverse realizations of the noise term; meanwhile, at any given cosmic time, sampling the ensemble of patches will originate a nontrivial spatial distribution of the various cosmological quantities. Finally, the overall behavior of the Universe will be obtained by averaging over the patch ensemble. We have assumed a very simple and physically reasonable parameterization of the noise term, gauging it against a wealth of cosmological datasets in the local and high-redshift Universe, including SN+Cepheid cosmography, baryon acoustic oscillations, cosmic chronometers, CMB first angular peak position, and age estimate from globular clusters.

We have found that, with respect to standard $\Lambda$CDM, the cosmic dynamics in the $\eta$CDM model is substantially altered by the noise in three main respects. First, an accelerated expansion is enforced at late cosmic times without the need for any additional exotic component (e.g., dark energy); the physical interpretation of this effect is that the overall ensemble of patches tends to drift toward an evolution dominated by low-density regions. Second, the global spatial curvature can stay small even in a low-density Universe constituted solely by matter and radiation. Third, matter can acquire an effective negative pressure at late times. We have also pointed our that the $\eta$CDM model is Hubble-tension free, meaning that the estimate of the Hubble constant from early and late-time measurements do not show marked disagreement as in $\Lambda$CDM. We have then provided specific predictions for the variance of the cosmological quantities induced by the noise at late cosmic times, which is found to be associated with the residual deviation from homogeneity/isotropy on large scales of order tens Mpcs. These could be tested with observations covering wide areas and large redshift intervals via a tomographic analysis; such observations could be quite challenging though within the reach of future surveys like \textit{Euclid} or \textit{LSST}.

Remarkably, the $\eta$CDM model admits an attractor solution in the infinite future with very peculiar features: the Universe will expand exponentially $a\propto e^{h_\infty\, t}$ with $e-$folding time given by a limiting value of the Hubble parameter $h_\infty$; the matter energy density parameter will saturate to a constant non-null value $\Omega_\infty$, and will be characterized by an effective equation of state $w_\infty\approx -1$, i.e. it will behave like a cosmological constant; the curvature will goes to zero. The limiting values $h_\infty$ and $\Omega_\infty$ are found to be only slightly smaller than the present ones; this implies that the Universe will spend a very long (actually, infinite!) amount of time hovering around very similar values of the cosmological parameters, so resolving any cosmic coincidence issue even in an wide sense without strongly invoking anthropic considerations.

In a future perspective, it would be welcome to: explore whether different types of noise (e.g., beyond the white noise approximation) can alter the cosmic dynamics; investigate whether the small residual anisotropy/inhomogeneity on large scales can appreciably perturb the gravitational metric, and estimate how the related corrections can impact on the estimation of the noise and of the cosmological parameters; study the evolution of perturbations and gauge the model parameters via an extended analysis on the overall CMB power spectrum, on the integrated Sachs \& Wolfe effect, and on CMB lensing; compute the growth function for cosmic structure formation, the effects on weak lensing probes, and hopefully address the $S_8$ tension; provide specific predictions for the observability of the implied anisotropies/inhomogeneities in future surveys, both via standard messengers and via gravitational waves.

We conclude with disclaiming that the $\eta$CDM framework presented here is just a very basic model, and that admittedly substantial work will be required for elevating it to a self-contained cosmology and for fully testing it against next-generation datasets. However, we very much hope that the new perspectives offered by the $\eta$CDM framework will contribute to trigger further attempts of explaining the observed phenomenology in the late-time cosmic expansion and of curing the plagues in the standard cosmological model, not necessarily invoking exotic form of energies or substantially revisiting of the standard (and up to now, observationally undefeated) Einstein's theory of gravity.

\begin{acknowledgements}
We thank the referee for a constructive report, and for the very useful suggestions on how to substantially improve the presentation of our work.
We acknowledge A. Bressan, G. Gandolfi, C. Ranucci and the GOThA team at SISSA for illuminating comments. AL dedicates the present work to M. Massardi, who has supported him in this great endeavor and tolerated long (and likely annoying to a radio astronomer like her) discussions on the subject. This work is funded by: the PRIN MIUR 2017 prot. 20173ML3WW, 'Opening the ALMA window on the cosmic evolution of gas, stars and supermassive black holes'; the EU H2020-MSCA-ITN-2019 Project 860744 'BiD4BESt: Big Data applications for black hole Evolution STudies'; the Fondazione ICSC - Spoke 3 Astrophysics and Cosmos Observations - National Recovery and Resilience Plan Project ID CN-00000013 'Italian Research Center on High-Performance Computing, Big Data and Quantum Computing' - Next Generation EU; the project 'Data Science methods for MultiMessenger Astrophysics \& Multi-Survey Cosmology' funded by the Italian Ministry of University and Research, Programmazione triennale 2021/2023 (DM n.2503 dd. 09/12/2019), Programma Congiunto Scuole; the INAF Large Grant 2022 funding scheme with the project 'MeerKAT and LOFAR Team up: a Unique Radio Window on Galaxy/AGN co-Evolution'.
\end{acknowledgements}

\begin{appendix}

\section{Stochastic differential equations}\label{sec|app_stocha}

In this Appendix we provide a primer on stochastic differential equations, pointing out the mathematical meaning of the noise term, the implications of adopting different stochastic prescriptions (e.g., Ito vs. Stratonovich) and the related origin of noise-induced drift terms; for more details and mathematical proofs the reader may have a look at the classic textbooks by Risken (1996) and by Paul \& Baschnagel (2013).

Consider a $n-$dimensional variable ${\bf x}=\{x_i\,;\, i=1\ldots n\}$ satisfying the stochastic differential equation
\begin{equation}\label{eq|sde_app}
\dot x_i = f_i({\bf x})+g_{ij}({\bf x})\, \eta_j(t)~,
\end{equation}
where $\{f_i\,;\, i=1\ldots n\}$ is called drift vector, $\{g_{ij}\,;\, i=1\ldots n\,,\, j=1\ldots m\}$ is called diffusion matrix, and $\{\eta_j\,;\, j=1\ldots m\}$ is a vector of independent noises with properties $\langle\eta_i(\tau)\rangle=0$ and $\langle\eta_i(\tau)\eta_j(\tau')\rangle=2\,\delta_{ij}\,\delta(\tau-\tau')$; boldface characters indicate vectors and summation over repeated indices is implicitly understood.

\begin{figure}[!t]
\figurenum{A1}
\centering
\includegraphics[width=\textwidth]{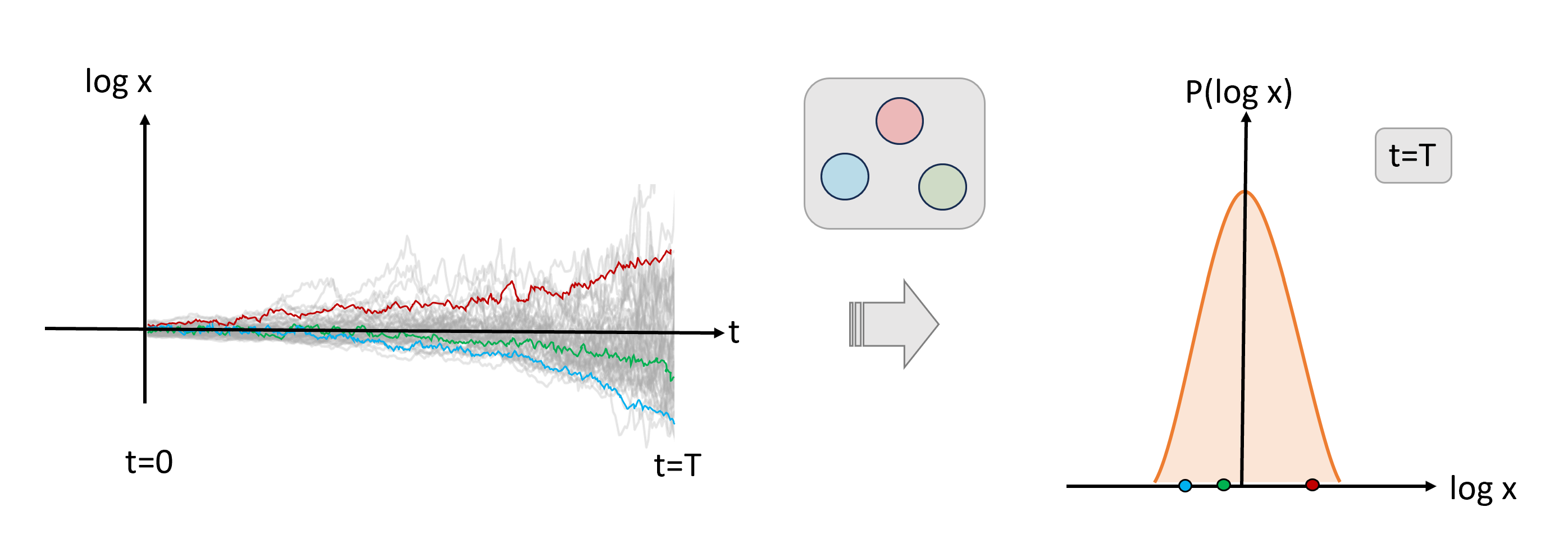}
\caption{Schematics that illustrates the mathematical meaning of a stochastic differential equation $\dot x = f(x)+g(x)\,\eta(t)$. Each realization of the noise $\eta(t)$ yields a different random walks (e.g., with reference in the main text, the walkers colored in red, blue and green correspond to the evolution of the density in three regions at different spatial locations). At given time $t=T$, sampling the value of the variable $x$ from the ensemble of walkers (e.g., from the different patches in the Universe) would yield a nontrivial probability distribution $P(x,T)$ of the variable $x$, which is the solution to the stochastic differential equation. The average $\bar{x}$ of this distribution will evolve in time as prescribed by Eqs. \ref{eq|ave_app}.}\label{fig|stochosmo}
\end{figure}

The mathematical meaning of the noise term $\eta_i(t)$ in the stochastic differential equation can be clarified as follows. For the sake of simplicity the reader may focus on the one-dimensional case, i.e. to the scalar equation $\dot x = f(x)+g(x)\, \eta(t)$, and may consider that $x$ refers to some physical property, e.g. with reference to the main text it could be the density $\rho$ associated to a given patch of the Universe. At every time $t$ the noise $\eta(t)$ should be considered as a value randomly extracted from a Gaussian distribution with zero mean and variance $2$ (by the conventional correlation property of the noise, but actually this can be put to any value by appropriately redefining $g$). For any realization of the noise, the variable $x(t)$ will execute a random walk, as schematically depicted in Fig. \ref{fig|stochosmo}; e.g., in the main text this can represent the specific evolution of the density in a given patch of the Universe. A different realization of the noise will yield a different walk, and so on so forth. At any given time $t=T$, sampling the value of the variable $x$ from the ensemble of walkers would yield a nontrivial probability distribution $\mathcal{P}(x,T)$ of the variable $x$, which is the solution to the stochastic differential equation; e.g., in the main text this distribution represents the spatial distribution of the density in different patches of the Universe.

Generally, solving a stochastic differential equation and obtaining the distribution $\mathcal{P}(x,T)$ requires a numerical approach (e.g., see textbook by Kloeden \& Platen 1992). Nevertheless, it is worth mentioning a few very basic examples with constant drift and diffusion coefficients and initial condition $x(0)=x_{\rm in}$, which admit a closed-form analytic solution. One is the Brownian motion, which describes a basic fluctuating process defined by the equation $\dot x = \mu +\sigma\, \eta(t)$; the solution is a normal distribution $\mathcal{P}(x,T)=\mathcal{N}(\bar{x},\sigma_x)$ with mean $\bar{x}=x_{\rm in}+\mu\, T$ and variance $\sigma_x^2=\sigma^2\,T$. Another one is the Ohrnstein-Uhlenbeck model, which is defined by $\dot x=\kappa\,(\mu-x)+\sigma\,\eta(t)$ and describes a process naturally falling back to an equilibrium level; the solution is a normal distribution $\mathcal{P}(x,T)=\mathcal{N}(\bar{x},\sigma_x)$ with mean $\bar{x} = \mu-(\mu-x_{\rm in})\, e^{-\kappa\,t}$ and variance $\sigma_x^2=(\sigma^2/2\kappa)\,(1-e^{-2\kappa\,t})$. Yet another is the geometric Brownian motion, which is defined by $\dot x = \mu\, x(t) +\sigma\,x(t)\,\eta(t)$ and is often exploited in finance to model stock prices; the solution is a log-normal distribution $\mathcal{P}( x,T)=\mathrm{Log}\mathcal{N}(\bar{x},\sigma_x^2)$ with average $\bar{x}  = x_{\rm in}\,e^{\mu\, t}$ and variance $\sigma_x^2=x_{\rm in}^2\,e^{2\mu\,t}\,(e^{\sigma^2\,t}-1)$.

In the general case of Eq.~(\ref{eq|sde_app}) the stochastic equations are vectorial and the drift/diffusion coefficients are nontrivial function of $\textbf{x}$, hence things get very complicated. When naively trying to solve the equations by integrating both sides in time, a well-known problem arises: the delta-correlated nature of the noise requires to give a meaning to the ill-defined integral
\begin{equation}
\int{\rm d}t \, g_{ij}[{\bf x}(t)]\,\eta_j(t)=\lim_{n\rightarrow \infty}\,\sum_{\ell=1}^n\,g_{ij}[{\bf x}(\tau_\ell)]\,[\eta_j(t_{\ell+1})-\eta_j(t_\ell)]~;
\end{equation}
note that here since the variables are stochastic, the limit is meant in the mean-square sense, i.e. $\lim_{n\rightarrow\infty} X_n=X$ stands for $\lim_{n\rightarrow\infty} \langle(X_n-X)^2\rangle =0$ where $\langle\cdot\rangle$ denotes an average over the ensemble.
For a smooth function the limit converges to a unique value independent of $\tau_\ell$, but this is not the case for the white noise $\eta$ since it fluctuates an infinite number of times with infinite variance in any small time interval (it is nowhere differentiable!). Thus the result depends on the choice of $\tau_\ell$ in the expression $g_{ij}[{\bf x}(\tau_\ell)]=g_{ij}[(1-\omega)\,{\bf x}(t_\ell)+\omega\,{\bf x}(t_{\ell+1})]$, with $\omega\in [0,1]$. Different choices of $\omega$ imply different rules for stochastic calculus; for example, for an arbitrary function $F[{\bf x}(t)]$ of the stochastic variables one can demonstrate that
\begin{equation}\label{eq|stochcalc}
{\rm d}_t F[{\bf x}(t)] = \partial_i\, F[{\bf x}(t)]\, \dot x_i + (1-2\omega)\,g_{ik}({\bf x})\,g_{jk}({\bf x})\,\partial_i\partial_j\, F[{\bf x}(t)]~
\end{equation}
holds.  Two common choices are the mid-point Stratonovich prescription ($\omega=1/2$) and the Ito pre-point prescription ($\omega=0$): the former is mostly used by physicists and the latter by mathematicians and numerical analysts. Note that the choice of the prescription is made on the basis of convenience, since plainly the overall dynamics of the system must be unique; in fact, it is possible to represent the same stochastic process in two arbitrary different prescriptions $\omega_1\rightarrow \omega_2$ just by appropriately rescaling the drift vector (e.g., Moreno et al. 2019)
\begin{equation}\label{eq|conv_app}
f_i({\bf x})\rightarrow f_i({\bf x})+2\,(\omega_1-\omega_2)\, g_{kj}({\bf x})\,\partial_k\,g_{ij}({\bf x})~.
\end{equation}

In the main text we adopt the Stratonovich prescriptions ($\omega=1/2$) since it has the advantage that, as shown by Eq.~(\ref{eq|stochcalc}),
the standard rules of calculus continue to hold and this makes easier to manipulate analytically the stochastic equations. However, this comes at a price: the average evolution $\bar{\bf x}(t)$ of the stochastic process in Eq.~(\ref{eq|sde_app}) is ruled by the equation
\begin{equation}\label{eq|ave_app}
\dot{\bar{x}}_i = f_i({\bf{\bar x}})+g_{kj}({\bf{\bar{x}}})\,\partial_k\,g_{ij}({\bf\bar{x}})~,
\end{equation}
hence does not only depend on the true drift $f_i$ (as it is true for the Ito prescription) but also on an additional noise-induced term (see Eq.~\ref{eq|conv_app}), that needs to be calculated. Demonstrating the above equation requires to evaluate the Kramers-Moyal coefficient and to derive the Fokker-Planck equation associated to the stochastic differential equations; a primer palatable for astrophysicists and cosmologists can be found in the Appendix A of Lapi \& Danese (2020) and in the classic textbook by Risken (1996). However, naively the
nature of the noise-induced drift can be understood by looking at the basic scalar equation $\dot x = f(x)+g(x)\, \eta(t)$. Taking the average on both sides yields
$\dot x = \langle f(x)\rangle +\langle g(x)\, \eta(t)\rangle$. Clearly
$\langle f(x)\rangle =f(\bar{x})$ holds since $f(x)$ is a purely deterministic function of $x$. The other term is less trivial, although at first sight one may think it is null since $\langle\eta(t)\rangle=0$ holds by the property of the white noise; however, this is not true since
as $\eta(t)$ varies also $x(t)$ and hence $g(x(t))$ changes, making $\langle g(x)\, \eta(t)\rangle$ finite; it is this instance that eventually leads to the noise-induced drift.

To make contact with the main text, note that the fundamental Eqs.~(\ref{eq|friednormrand}) constitute a particular case of Eq.~(\ref{eq|sde_app}) for the three cosmological variables ${\bf x}\equiv (h,\Omega_m,\Omega_\gamma)$ when just an independent noise $\eta$ is present. Specifically,
one can define the following components: for the variable vector
$x_0=h$, $x_1=\Omega_m$, $x_2=\Omega_\gamma$; for the drift vector
$f_0=h^2\,(-1-\Omega_m/2-\Omega_\gamma)$, $f_1=\Omega_m\,h\,(-1+\Omega_m+2\Omega_\gamma)$, $f_2=\Omega_\gamma\,h\,(-2+\Omega_m+2\Omega_\gamma)$;
for the noise vector $\eta_0=\eta$ and $\eta_1=\eta_2=0$; for the diffusion matrix
$g_{00}=\zeta/2\,(\Omega_m+\Omega_\gamma)\,h^{\alpha+1}$, $g_{10}=\zeta\,(1-\Omega_m-\Omega_\gamma)\,\Omega_m\,h^\alpha$,
$g_{20}=\zeta\,(1-\Omega_m-\Omega_\gamma)\,\Omega_\gamma\,h^\alpha$, with all the other components null. Then it is simple algebra to verify that Eqs.~(\ref{eq|friednormrand}) can be put in the vectorial form expressed by Eq.~(\ref{eq|sde_app}).

For the system of Eqs.~(\ref{eq|friednormrand}) computing the noise-induced drift and so obtaining the average evolution described by Eqs.~(\ref{eq|friednew}) is straightforward but quite tedious. For the reader convenience we report some details on the computation for the component $x_0=h$; considering only the non-null elements in the summation involved in Eq.~(\ref{eq|ave_app})  one has  $\dot{\bar{x_0}}=f_0(\bar{x})+g_{00}(\bar{x})\,\partial_{x_0}\,g_{00}(\bar{x})+g_{10}(\bar{x})\,
\partial_{x_1}\,g_{00}(\bar{x})+g_{20}(\bar{x})\,\partial_{x_2}\,g_{00}(\bar{x})$
or in more explicit form $\dot{\bar{h}} = \bar{h}^2\,(-1-\bar{\Omega}_m/2-\bar{\Omega}_\gamma)+(\zeta^2/2)\,\bar{\Omega}_m\, (1-\bar{\Omega}_m-\bar{\Omega}_\gamma)\,\bar{h}^{2\alpha+1}+(\zeta^2/2)\,\bar{\Omega}_\gamma\, (1-\bar{\Omega}_m-\bar{\Omega}_\gamma)\,\bar{h}^{2\alpha+1}+
(\zeta^2/4)\,(\alpha+1)\,(\bar{\Omega}_m+\bar{\Omega}_\gamma)^2\,\bar{h}^{2\alpha+1}$. Simplifying the last three addenda one finally obtains
$\dot{\bar{h}} = \bar{h}^2\,(-1-\bar{\Omega}_m/2-\bar{\Omega}_\gamma)+(\zeta^2/2)\,(\bar{\Omega}_m+\bar{\Omega}_\gamma)\, [1-(1-\alpha)\,(\bar{\Omega}_m+\bar{\Omega}_\gamma)/2]\,\bar{h}^{2\alpha+1}$ which is the first of Eqs.~(\ref{eq|friednew}); this explicitly includes on the right hand side both the true and the noise induced drifts. The average evolution equations for $\Omega_m$ and $\Omega_\gamma$ can be derived analogously.

\section{Diffusion bridges}\label{sec|app_bridge}

In this Appendix we provide some basic information on diffusion bridges, i.e. stochastic processes pinned at both ends in some values. For more details the reader may consult the papers by Pedersen (1995), Durham \& Gallant (2002), Delyon \& Hu (2006), Lindstrom (2012), Bladt \& Sorensen (2014), Whitaker et al. (2017), Heng et al. (2022).

Suppose to have a $N-$dimensional continuous stochastic process ${\bf x}(t)=\{x_i\,;\, i=1\ldots N\}$ for $t\in [0,T]$ satisfying the system $\dot x_i(t)=f_i({\bf x},t)+g_{ij}({\bf x},t)\, \eta_j(t)$ in the Stratonovich sense, with boundary value ${\bf x}(T)={\bf x}_T$ and with the property that the random term becomes negligible at early times; here $\{f_i\,;\, i=1\ldots N\}$ is a drift vector, $\{g_{ij}\,;\, i,j=1\ldots N\}$ is a diffusion matrix, and $\{\eta_j\,;\, j=1\ldots N\}$ a vector of independent noises; repeated summation convention is adopted.

To (approximately) solve the stochastic system, the idea is to partition the process ${\bf x}={\bf \bar x}+{\bf \tilde x}$ in an average ${\bf \bar x}$ and in a residual random ${\bf \tilde x}$ component, satisfying the system of coupled equations:
\begin{equation}\label{eq|diffbridge}
\left\{
\begin{aligned}
\dot{\bar{x}}_i(t) &= f_i({\bf \bar{x}})+g_{kj}({\bf \bar{x}})\, \partial_k\,g_{ij}({\bf \bar{x}})~~, & ~{\bf \bar{x}}(T)={\bf x}_T\\
\\
\dot{\tilde{x}}_i(t) &= - \cfrac{\tilde{x}_i(t)}{T-t} + g_{ij}({\bf x})\, \eta_j(t)~~, & ~ {\bf \tilde{x}}(0)={\bf 0}~.
\end{aligned}
\right.
\end{equation}
The first equation describes the average evolution of the system and is analogous to Eq.~(\ref{eq|ave_app}): the first term on the right hand side is the true drift, while the second term is the noise-induced one. Being an ordinary differential equation, this can be evolved backward to the initial time via standard methods, to provide an initial condition ${\bf x}(0)={\bf x}_0$ for the full system.

The second equation is less trivial: it describes a residual random process (in the Ito sense), executing a conditioned diffusion that starts in ${\bf \tilde{x}}(0)={\bf 0}$ by the boundary condition, and is also forced to end in ${\bf \tilde{x}}(T)={\bf 0}$ by the spurious drift term $-\tilde{x}_i/(T-t)$, in such a way that the overall stochastic process ${\bf x}={\bf \bar x}+{\bf \tilde x}$ behaves as expected, with ${\bf x}(0)={\bf x}_0$ and ${\bf x}(T)={\bf x}_T$ at the extremes. The origin of the spurious drift is a bit technical. In fact, it can be can be demonstrated rigorously (e.g., Rogers \& Williams 2000) that a diffusion process $\dot{x}_i(t) = g_{ij}({\bf x},t)\, \eta_j(t)$ with null drift constrained to start and end in zero (in the Ito convention) is equivalent to an unconditioned process with a spurious drift that forces the walker to hit the final condition, of the form $\dot{x}_i(t) = g_{ik}({\bf x},t)\,g_{kj}({\bf x},t)\,\nabla_j\,\mathcal{P}({\bf 0},T|{\bf x},t)+g_{ij}({\bf x},t)\, \eta_j(t)$ where $\mathcal{P}$ is the transition density of the unconditioned process. Since $\mathcal{P}$ in most cases is untractable, a linear Gaussian approximation to it is often adopted (the so called modified diffusion bridge, see Durham \& Gallant 2002; Delyon \& Hu 2006) that leads to the spurious drift reported in the second of Eqs.~(\ref{eq|diffbridge}). Involving an ordinary and a stochastic differential equation, the above system can be solved forward in time via standard numerical techniques, with the only caveat that the argument of the diffusion matrix on the right hand side of the second equation is the full process ${\bf x}=\bar{{\bf x}}+\tilde{{\bf x}}$, hence the second equation requires the solution of the first as an input.

In the main text, all the above has been applied to the original stochastic system Eqs.~(\ref{eq|friednormrand}), to obtain Eqs.~(\ref{eq|friednew}) for the average evolution exploited in Sect. \ref{sec|deterministic} (see derivation in Appendix \ref{sec|app_stocha}), and  Eqs.~(\ref{eq|stocha}) for the random part exploited in Sect. \ref{sec|stochasticity}. For example, to derive the latter,
consider the zero-th component of the second Eqs.~(\ref{eq|diffbridge}) that reads
$\dot{\tilde{x}}_0(t) = - \tilde{x}_0(t)/(T-t) + g_{00}({\bf x})\, \eta(t)$.
Recalling from Appendix \ref{sec|app_stocha} that ${\bf x}\equiv (h,\Omega_m,\Omega_\gamma)$ and that $g_{00}=(\zeta/2)\,(\Omega_m+\Omega_\gamma)\,h^{\alpha+1}$ in terms of the cosmological variables, we obtain
$\dot{\tilde{h}}(t) = - \tilde{h}/(T-t) + (\zeta/2)\,(\Omega_m+\Omega_\gamma)\,h^{\alpha+1}\, \eta(t)$, which is the first of Eqs.~(\ref{eq|stocha}). The equations for the other components corresponding to $\Omega_m$ and $\Omega_\gamma$ are derived analogously.

\section{Validation of the fitting pipeline}

In this appendix we validate our fitting pipeline described in Sect. \ref{sec|cosmofit}, by applying it to the
(curvature-free) $\Lambda$CDM model. To this purpose, we solve the evolution equations
\begin{equation}\label{eq|LCDM}
\left\{
\begin{aligned}
\dot{h} &= h^2\,\left(-1-\cfrac{\Omega_m}{2}-\Omega_\gamma-\Omega_\Lambda\cfrac{1+3\,w_\Lambda}{2}\right)\\
\\
\dot{\Omega}_m &=\Omega_m\, h\, [-1+\Omega_m+2\,\Omega_\gamma+\Omega_\Lambda\,(1+ 3\,w_\Lambda)]\\
\\
\dot{\Omega}_\gamma &=\Omega_\gamma\, h\, [-2+\Omega_m+2\,\Omega_\gamma+\Omega_\Lambda\,(1+3\,w_\Lambda)]~,\\
\\
\dot{\Omega}_\Lambda &=\Omega_\Lambda\, h\, [-1+\Omega_m+2\,\Omega_\gamma+\Omega_\Lambda\,(1+3\,w_\Lambda)-3\,w_\Lambda]~,\\
\end{aligned}
\right.
\end{equation}
with boundary conditions $(h_0, \Omega_{m,0}, \Omega_{\gamma,0},\Omega_{\Lambda,0})$. For the sake of simplicity, as in the main text we set $\Omega_{\gamma,0}\,h_0^2\approx 2.47\times 10^{-5}$ and $\Omega_{b,0}\, h_0^2\approx 0.0222$; we also consider only the case of constant equation of state $w_\Lambda=-1$ for dark energy. We then fit such a model to the cosmological datasets described in Sect. \ref{sec|cosmofit} and perform Bayesian inference on the normalized Hubble constant $h_0$, the present energy density parameters of matter $\Omega_{m,0}$ and of dark energy $\Omega_{\Lambda,0}$.

\begin{figure}[!t]
\figurenum{C1}
\centering
\includegraphics[width=0.8\textwidth]{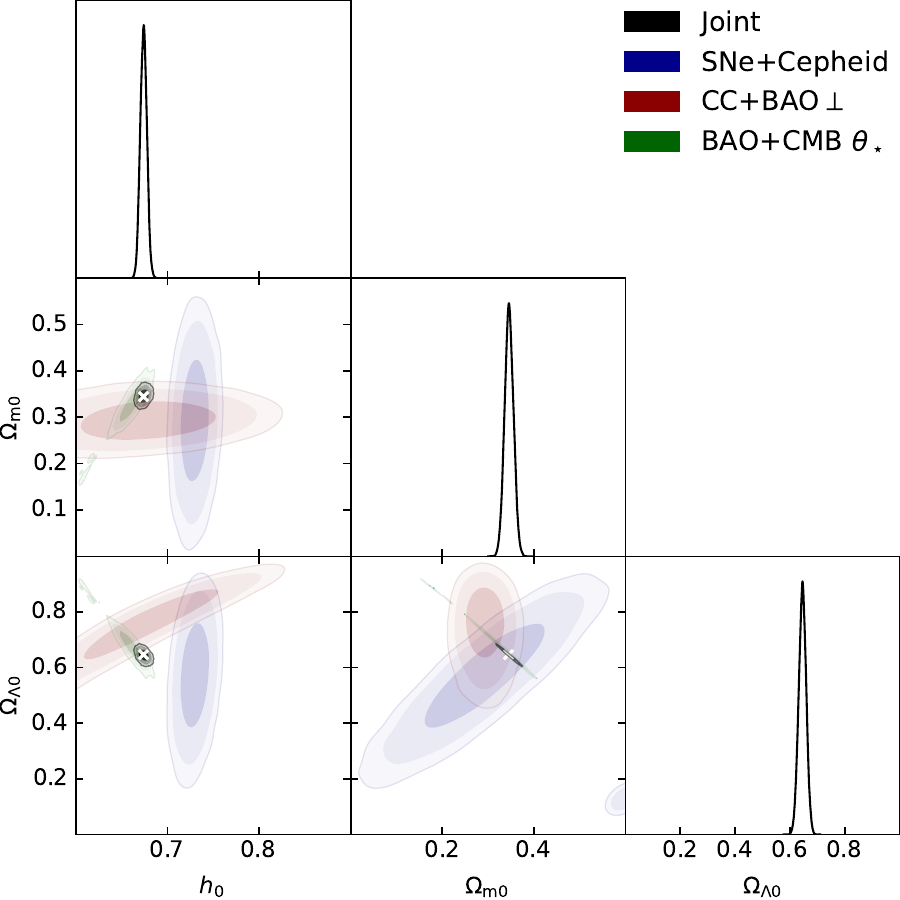}
\caption{MCMC posterior distributions in the (curvature-free) $\Lambda$CDM model, for the normalized hubble constant $h_0$,
the matter density parameter $\Omega_{m,0}$, and the dark energy density parameter $\Omega_{\Lambda,0}$. Colored contours/lines refer to different observables: blue for SN + Cepheid, orange for CC + transverse BAOs, green for isotropic BAO + CMB first peak angular position, and black for joint. The contours show $1-2-3\sigma$ confidence intervals, crosses mark the maximum likelihood estimates of the joint analysis, and the marginalized distributions of the joint analysis are reported on the diagonal panels in arbitrary units (normalized to 1 at their maximum value).}\label{fig|MCMC_LCDM}
\end{figure}

The marginalized constraints are shown in Fig. \ref{fig|MCMC_LCDM} and reported in Table \ref{Table|MCMC_LCDM}. The fits to the individual datasets produce the expected results, with SN+Cepheid mainly setting $h_0$ and constraining with some degeneracy $\Omega_{m,0}$ and $\Omega_{\Lambda,0}$; the degeneracy is removed by BAO+CMB data, which basically requires a flat Universe. From the one-dimensional posterior it is also quite evident the $h_0$ tension between late-time and early-time measurements, with BAO+CMB data preferring lower values with respect to SN+Cepheid. The $H_0$ tension is at the origin of the weird behavior of the joint analysis posterior, which tends to be maximized in a region where the Universe is flat but close to the $h_0$ determination from CMB+BAO; clearly the joint analysis is barely significant when the individual fits are discordant on one crucial parameter like in the $\Lambda$CDM model.

\begin{deluxetable*}{lccccccccccccccccccccccccc}\label{Table|MCMC_LCDM}
\tablenum{A1}
\tablecaption{Marginalized posterior estimates in terms of mean and $1\sigma$ confidence interval [and bestfit value] for the fits with the (curvature-free) $\Lambda$CDM model to different cosmological datasets, as listed in the first column. Other columns report the values of the normalized Hubble constant $h_0$, present matter energy density $\Omega_{m,0}$, dark energy density $\Omega_{\Lambda,0}$, and the reduced $\chi_r^2$ of the various fits.}
\tablewidth{0pt}
\tablehead{\colhead{Dataset} & & \colhead{$h_0$} & \colhead{$\Omega_{m,0}$} & \colhead{$\Omega_{\Lambda,0}$} & \colhead{$\chi^2_r$}}
\startdata
\\
\\
Joint & &$0.674^{+0.004}_{-0.004}$ [0.67] & $0.35^{+0.01}_{-0.01}$ [0.35] & $0.65^{+0.01}_{-0.01}$ [0.65] & 0.74\\
\\
SNe+Cepheid & &$0.734^{+0.005}_{-0.015}$ [0.73] & $0.30^{+0.08}_{-0.10}$ [0.29] & $0.56^{+0.14}_{-0.11}$ [0.57] & 0.29\\
\\
CC+BAO$_\perp$ & &$0.69^{+0.04}_{-0.06}$ [0.69] & $0.31^{+0.01}_{-0.05}$ [0.29] & $0.75^{+0.10}_{-0.06}$ [0.78] & 0.55\\
\\
BAO+CMB & &$0.66^{+0.01}_{-0.01}$ [0.66] & $0.33^{+0.03}_{-0.02}$ [0.32] & $0.68^{+0.03}_{-0.04}$ [0.68] & 0.29\\
\\
\\
\enddata
\end{deluxetable*}

\end{appendix}

\end{document}